\documentclass[12pt]{article}

\usepackage{latexsym,amsmath,amssymb,amsfonts,amsthm,bbm}
\usepackage{natbib}
\usepackage[table]{xcolor} 
\usepackage[colorlinks=true,citecolor=blue]{hyperref}
\usepackage{enumerate}
\usepackage{subfig}
\usepackage{graphicx}
\usepackage{graphics,enumerate}
\usepackage{fancybox}
\usepackage{algorithmicx}
\usepackage[noend]{algpseudocode}
\usepackage{mathrsfs}
\usepackage{bm}
\usepackage[lined,boxed,commentsnumbered]{algorithm2e}
\usepackage{tikz}
\usetikzlibrary{shapes,snakes}
\usetikzlibrary{matrix}
\usepackage{multirow}
\usepackage{booktabs}
\usepackage{diagbox}
\newtheorem{theorem}{Theorem}

\newtheorem{corollary}{Corollary}

\newtheorem{lemma}{Lemma}
\newtheorem{definition}{Definition}
\newtheorem{rmk}{Remark}
\numberwithin{equation}{section}
\numberwithin{figure}{section}
\numberwithin{table}{section}
\numberwithin{theorem}{section}
\numberwithin{lemma}{section}
\numberwithin{corollary}{section}

\def\subSNR{_{\scriptscriptstyle \sf SNR}}
\usepackage{chngcntr}
\usepackage{apptools}

\newenvironment{eq}{\begin{equation}\begin{aligned}}{\end{aligned}\end{equation}}
\newenvironment{eq*}{\begin{equation*}\begin{aligned}}{\end{aligned}\end{equation*}}

\def\trans{^{\scriptscriptstyle \sf T}}
\def \sE {{\sf E}}
\def \sV {{\sf Var}}

\def \one {\mathbf{1}}

\def \Pscr {\mathscr{P}}
\def \Gcal {\mathbb{G}}

\def \Wscr {\mathscr{W}}

\def \Hcal {\mathcal{H}}
\def \Mcal {\mathcal{M}}
\def \Nscr {\mathscr{N}}

\def \Fcal {\mathcal{F}}
\def \Vcal {\mathcal{V}}

\def\diag {\mathop{\rm diag}}
\def\argmin{\mathop{\rm argmin}}

\def\Abb{\mathbb{A}}
\def\Bbb{\mathbb{B}}
\def\Obb{\mathbb{O}}
\def\Wbb{\mathbb{W}}

\def\Zscr{\mathscr{Z}}

\newcommand{\norm}[1]{\left\lVert#1\right\rVert}

\usepackage{mathtools}

\makeatletter
\newcommand{\distas}[1]{\mathbin{\overset{#1}{\kern\z@\sim}}}%
\newsavebox{\mybox}\newsavebox{\mysim}
\newcommand{\distras}[1]{%
	\savebox{\mybox}{\hbox{\kern3pt$\scriptstyle#1$\kern3pt}}%
	\savebox{\mysim}{\hbox{$\sim$}}%
	\mathbin{\overset{#1}{\kern\z@\resizebox{\wd\mybox}{\ht\mysim}{$\sim$}}}%
}
\makeatother

\definecolor{darkred}{RGB}{150,50,50}
\definecolor{brown}{RGB}{250,100,100}
\definecolor{green}{RGB}{000,150,100}
\definecolor{purple}{RGB}{250,000,180}

\def\Zbb{\mathbb{Z}}
\def\Zbbhat{\widehat{\Zbb}}
\def\Abbhat{\widehat{\Abb}}

\def\Ebb{\mathbb{E}}
\def\Ubb{\mathbb{U}}
\def\Ubbhat{\widehat{\Ubb}}
\def\Dbb{\mathbb{D}}

\def\Vbb{\mathbb{V}}
\def\Omegatilde{\widetilde{\Omega}}
\def\Rcal{\mathcal{R}}
\def\blambda{\boldsymbol{\lambda}}
\def\bh{\boldsymbol{h}}
\def\ba{\boldsymbol{a}}

\def\NMI{\mbox{NMI}}
\def\MS{\mbox{MSE}}

\usepackage{footnotebackref}
\usepackage{setspace}

\newenvironment{keywords}
{\bgroup\leftskip 20pt\rightskip 20pt \noindent{\bf Keywords:} }%
{\par\egroup\vskip 0.25ex}

\begin{document}
	\title{Multi-view Banded Spectral Clustering with Application to ICD9 Clustering}
	
	\author{Luwan Zhang $^1$, Katherine Liao$^2$, Isaac Kohane$^3$, Tianxi Cai$^1$ \\ 
	\em $^1$Department of Biostatistics, Harvard T.H. Chan School of Public Health, Boston MA \\
	\em $^2$Division of Rheumatology, Brigham and Women's Hospital, Boston MA \\
	\em $^3$Department of Biomedical Informatics, Harvard Medical School, Boston MA}
    \date{}
	
	\maketitle
	\begin{abstract}

    Despite recent development in methodology, community detection remains a challenging problem. Existing literature largely focuses on the standard setting where a network is learned using an observed adjacency matrix from a single data source. Constructing a shared network from multiple data sources is more challenging due to the heterogeneity across populations. Additionally, no existing method leverages the prior distance knowledge available in many domains to help the discovery of the network structure. To bridge this gap, in this paper we propose a novel spectral clustering method that optimally combines multiple data sources while leveraging the prior distance knowledge. The proposed method combines a banding step guided by the distance knowledge with a subsequent weighting step to maximize consensus across multiple sources. Its statistical performance is thoroughly studied under a multi-view stochastic block model. We also provide a simple yet optimal rule of choosing weights in practice.  The efficacy and robustness of the method is fully demonstrated through extensive simulations. Finally, we apply the method to cluster the International classification of diseases, ninth revision (ICD9), codes and yield a very insightful clustering structure by integrating information from a large claim database and two healthcare systems.

	\end{abstract}
	
	\vskip 10pt
	\begin{keywords}
		 multi-view, banding, spectral clustering, community detection, stochastic block model
	\end{keywords}

	\clearpage
	\newpage
	
	\baselineskip=24pt

\section{Introduction}

Collapsing interchangeable or highly similar entities into a single group is a problem of great importance and necessity in a wide range of areas.  It enables signal enhancement, dimension reduction and variable selection, while ensuring reproducibility and interpretability on subsequent research results. The concept of ``nearly equivalent" entities naturally arises in many fields presenting in various forms to serve for different research purposes. For example, to identify genetic variants that increase susceptibility to a disease (or other phenotype of interest), it can be nearby markers that are usually correlated and hence form a linkage disequilibrium (LD) block.  In natural language processing, it can be synonyms entitled with a similar meaning and occurring in a similar context. In phenome-wide association studies (PheWAS), it can be the International Classification of Disease (ICD) codes that essentially describe the same disease but differ in details such as the affected anatomical areas. Similarly, in brain image analysis, voxels in the same region with a common neurological function can also be viewed as interchangeable.


The problem of collapsing interchangeable entities can be translated into a statistical problem of community detection. More specifically, let $v_i$ refer to the $i$-th entity and $\Vcal=\{v_i\}_{i=1}^n$ represent the whole vertex set, the goal is to seek a partition such that $$\Vcal=\cup_k \Vcal_k, \Vcal_k \cap \Vcal_l = \emptyset, \forall k\neq l.$$ where $\Vcal_k$ denotes the $k$-th community and any entity pair $(v_i, v_j)$ from this group is stochastically equivalent.  
To infer such partition with a single observed similarity matrix $\Wbb$, many statistical methods have been proposed and witnessed huge success in numerous applications \cite[e.g.]{shi2000normalized, ng2002spectral, newman2006modularity, bickel2009nonparametric, zhao2012consistency}. 
Statistical properties of the clustering have also been established under the framework of stochastic block model (SBM) \citep{holland1983stochastic} and its extension to allow for degree heterogenity \citep[e.g.]{karrer2011stochastic,rohe2011spectral, qin2013regularized, lei2015consistency,jin2015fast}.  

When multiple similarity matrices from different data sources become available, an important task is to develop an effective synergistic integration strategy that leads to a better inference on the underlying network structure.  To this end, a series of multi-view clustering methods have been proposed \citep[e.g.]{blaschko2008correlational,chaudhuri2009multi,cai2011heterogeneous,kumar2011co,xia2014robust},  where a data source is also termed as a view.	
Despite their empirical success, little theoretical justifications have been provided until recently \cite{han2014consistent} and \cite{paul2016consistent} showed some consistency results. In addition, optimal approaches of combining all views to account for their heterogeneity remain elusive.

Another limitation of these existing methods is that they require input being unweighted, binary-valued similarity matrices. This would largely hamper the applicability to settings with similarity matrices being weighted and real-valued. In particular, real-valued measures of similarity are frequently used in numerous contemporary biological and clinical applications. For example, with the recent emergence of  \textsl{word2vec} algorithms \citep{mikolov2013efficient}, biological sequences (e.g. genes, proteins), clinical concepts describing disease conditions, and ICD codes have been represented by Euclidean vectors \citep[e.g.]{asgari2015continuous, nguyen2016control,choi2016learning,ng2017dna2vec}, with pairwise similarity frequently summarized by the real-valued cosine score. 

In addition to multi-view, prior knowledge on the distance between the nodes is often available through established ontologies,  especially in biomedical domains. Such distance information can potentially help the discovery of the network structure in that more distant nodes suggested by the ontology are less likely grouped together. For example, when nodes represent chromosomal loci, clusters of homogeneous genes tend to be adjacent genes. For brain graph connectivity, spatially further apart voxels are less likely considered as belonging to the same biological group. Hierarchical structures have also been curated for a wide range of clinical concepts. For example, the unified medical language system (UMLS) provides a relational database for clinical terms used in medical language \citep{humphreys1993umls}. Relationships between different disease phenotypes, medical concepts are described in the ICD hierarchy and the human phenotype ontology such as SNOMED-CT and MedDRA. Nodes further apart on these hierarchies less likely belong to the same group. 

To the best of our knowledge, no existing clustering method incorporates multiple, say $m$, real-valued similarity matrices $\{\Wbb^s,s=1,...,m\}$ or leverages prior information on the distance between the nodes.  In this paper, we propose a novel two-step muti-view banded spectral clustering (mvBSC) method to bridge this gap. The mvBSC method leverages the prior knowledge by restricting the parameter space to a class of decaying matrices and integrates information from $m$ sources by performing spectral clustering on a convex combination of $m$ membership-encoded matrices. Although the performance of clustering accuracy improves due to banding, our procedure is robust to the banding assumption and remains valid in the absence of this operation.

The rest of paper is organized as follows. In Section 2, we give a formal description of the multi-view stochastic block model and provide assumptions on the parameter space.  Section 3 details the proposed multi-view banded spectral clustering method and provides theoretical justifications under the multi-view stochastic block model. Simulations are given in Section 4 to demonstrate the efficacy and robustness of the proposed method. In Section 5, we apply the proposed method to the ICD9 coding system and yield a very insightful clustering structure by integrating information from a large claim database and two healthcare systems. Concluding remarks and discussions are given in Section 6.

	\section{Multi-view Stochastic Block Model}
	\subsection{Notations}
  For any matrix $\Abb \in \Rcal^{p \times p}$, let $\Abb_{i\cdot}$ and $\Abb_{\cdot i}$ respectively denote the $i$th row and column of $\Abb$, and let $\|\Abb\|_F$ and $\|\Abb\|$  respectively denote its Frobenius norm and  spectral norm. For any two matrices $\Abb, \Bbb \in \Rcal^{p \times p}$,  $\Abb \lesssim \Bbb$ means that $\Abb \leq c\Bbb$ for some constant $c > 0$ and $\Abb \gtrsim \Bbb$ means that $\Abb \geq c\Bbb$ for some constant $c > 0$. $ \Abb \asymp \Bbb$ is equivalent as $\Abb \lesssim \Bbb \lesssim \Abb$. For any vector $\ba = (a_1,...,a_p)\trans$, let $\diag(\ba)$ denote the corresponding $p \times p$ diagonal matrix with diagonals being $\ba$ and $\|\ba\|_2$ be its $\ell_2$-norm. Let $\one_p = (1,...,1) \trans$ denote the all-one vector in $\Rcal^p$ and $\ba \geq 0$ indicate the coordinatewise non-negativity. For $x, y \in \Rcal$, $x=o(y)$ means ${x \over y} = o(1)$. Let $I(\cdot)$ denote the indicator function.
	
    Suppose the data for analysis consist of $m$ similarity matrices, $\{\Wbb^s, s = 1, ..., m\}$, representing $m$ undirected weighted graphs, $\Gcal^s = \{\Vcal, \Wbb^s\}$, where $\Wbb^s = [W_{ij}^s]_{n \times n}$, $W_{ij}^s$ is
	the similarity between the nodes $v_i$ and $v_j$ based on the $s$th view, and we 
	assume that these $m$ graphs share the same vertex set $\Vcal=\{v_i\}_{i=1}^n$ which has a non-overlapping $K$-partition network structure
	$$\Vcal = \{\cup_{k=1}^K \Vcal_k, \Vcal_k \cap \Vcal_l = \emptyset, \forall 1\leq k <l \leq K\} , $$
	where $\Vcal_k = \{v_i\}_{i: g_i = k}$ and $g_i \in \{1, ..., K\}$ indexes which group $v_i$ belongs to.  Let $n_k := |\Vcal_k|$ denote the size of the $k$-th cluster.
	The partition can also be represented by a group membership matrix 
	$$\Zbb^* = [\Zbb^*_{ik}]_{n \times K} \in \Zscr_{n,K}, \quad \mbox{where}\quad \Zbb^*_{ik} =  I(v_i \in \Vcal_k) = I(g_i = k),$$
	and $\Zscr_{n,K}$ consists of all possible $K$-group membership matrices for $n$ nodes. Denote its associated class of projector matrices by
	$$\Pscr_{n,K} = \left\{ \Zbb \left[\diag(\one_n \trans \Zbb)\right]^{-1} \Zbb\trans: \Zbb \in \Zscr_{n,K} \right\} $$
	In particular, denote by $P_{\Zbb}$ the projector matrix $\Zbb \left[\diag(\one_n \trans \Zbb)\right]^{-1} \Zbb\trans$ that projects any $n$-dimensional vector to the $K$-dimensional subspace spanned by the columns of $\Zbb$. 
	Throughout, we assume that $K$ is known and remains as a constant for all theoretical analyses. Strategies for choosing $K$ in practice will be discussed in Section 5. 
	
	\subsection{Assumptions on mvSBM}
	We aim to optimally combine information from the m-views, $\{\Wbb^s = [W_{ij}^s]_{n \times n}, s = 1, ..., m\}$, to learn about the network structure through an mvSBM such that $\forall \; 1 \leq i < j \leq n$,
	\begin{align}
	\label{eq: model}
	 \sE_{\Zbb^*} W_{ij}^s  \equiv \Wscr_{ij}^s =  \Omega^s_{{g_i}{g_j}},  \quad \Wscr^s = [\Wscr_{ij}^s]_{i=1,...,n}^{j=1,...,n} = \Zbb^*\Omega^s\Zbb^{*\trans},  \quad |W_{ij}^s| \leq L, \quad \sV_{\Zbb^*} (W_{ij}^s ) = \sigma_s^2 ,		\end{align} 
	where $\Omega^s = [\Omega_{kl}^s]_{k=1,...,K}^{l = 1, ...,K} \in [-L, L]^{K \times K}$ is a symmetric and positive definite matrix of rank $K$, $L < \infty$ is a constant, $\sE_{\Zbb^*} $ and $\sV_{\Zbb^*}$ respectively denote the expectation and variance given the membership matrix $\Zbb^*$. 	Thus, under the mvSBM, the hidden membership matrix $\Zbb^*$ is shared across all m views, but the connection intensities encoded by $\Omega^s$ may vary. 
    For each diagonal element $W^s_{ii} $,  it can be either considered as a yet another independent bounded random variable following the model specified in (\ref{eq: model}), or it can be treated as a non-random constant $\omega_0 \in [-L,L]$. Without loss of generality, we consider hereafter the latter case , which includes most commonly used similarity measures.
   
	\begin{rmk}
		The traditional definition of a SBM given in \cite{holland1983stochastic} requires that each $\Wbb^s$ is 0/1 valued in that each off-diagonal entry corresponds to an independent Bernoulli random variable whose probability depends only on the block memberships of the two nodes.  Here we extend to a real-valued setting to allow for more generality. In fact, \cite{karrer2011stochastic} was the first attempt to extend the applicability of a SBM in which a Poisson model is imposed to allow for multiple edges between any two nodes. 
	\end{rmk}
	
\begin{figure}[htbp]
		\begin{center}
			\includegraphics[scale=0.5]{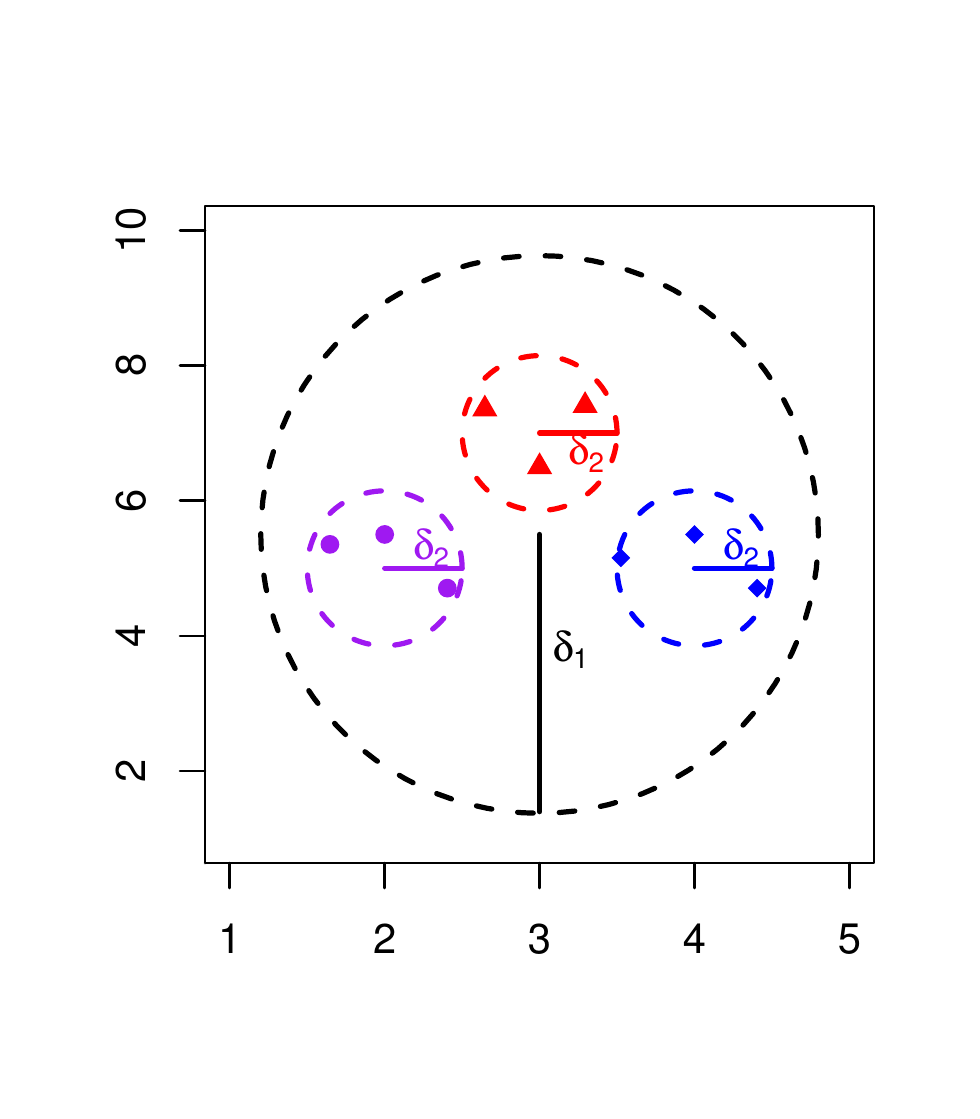}
			\caption{Effect of $\delta$: two different choices of $\delta$ lead to two different thresholding schemes. Nodes from the same cluster are in the same color and symbol.  }
			\label{fig: delta}
		\end{center}
	\end{figure}

\begin{figure}[htbp]
\begin{center}
	\subfloat[$\Omega^s$]{\includegraphics[width=5cm,height=5cm]{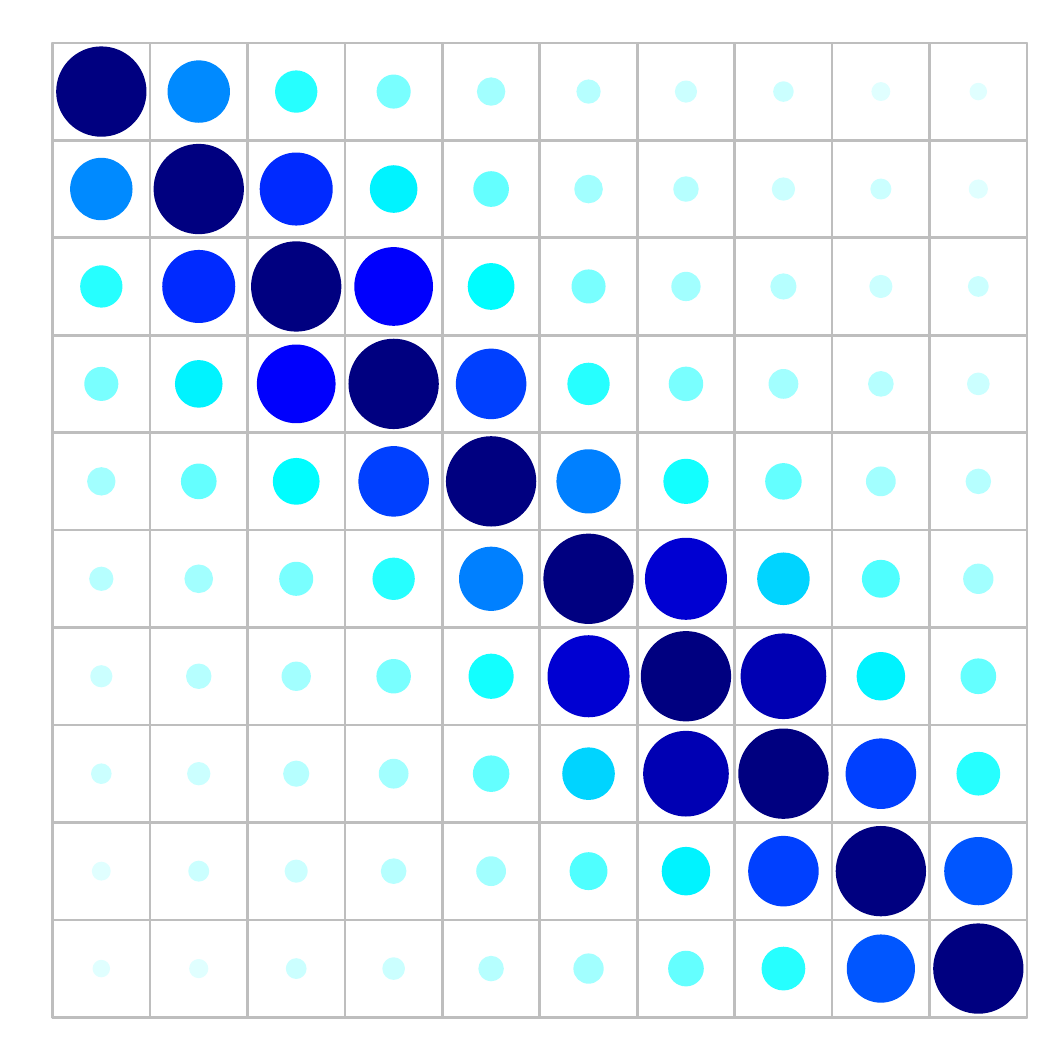}} \quad
		\subfloat[$\Wscr^s$]{\includegraphics[width=5cm,height=5cm]{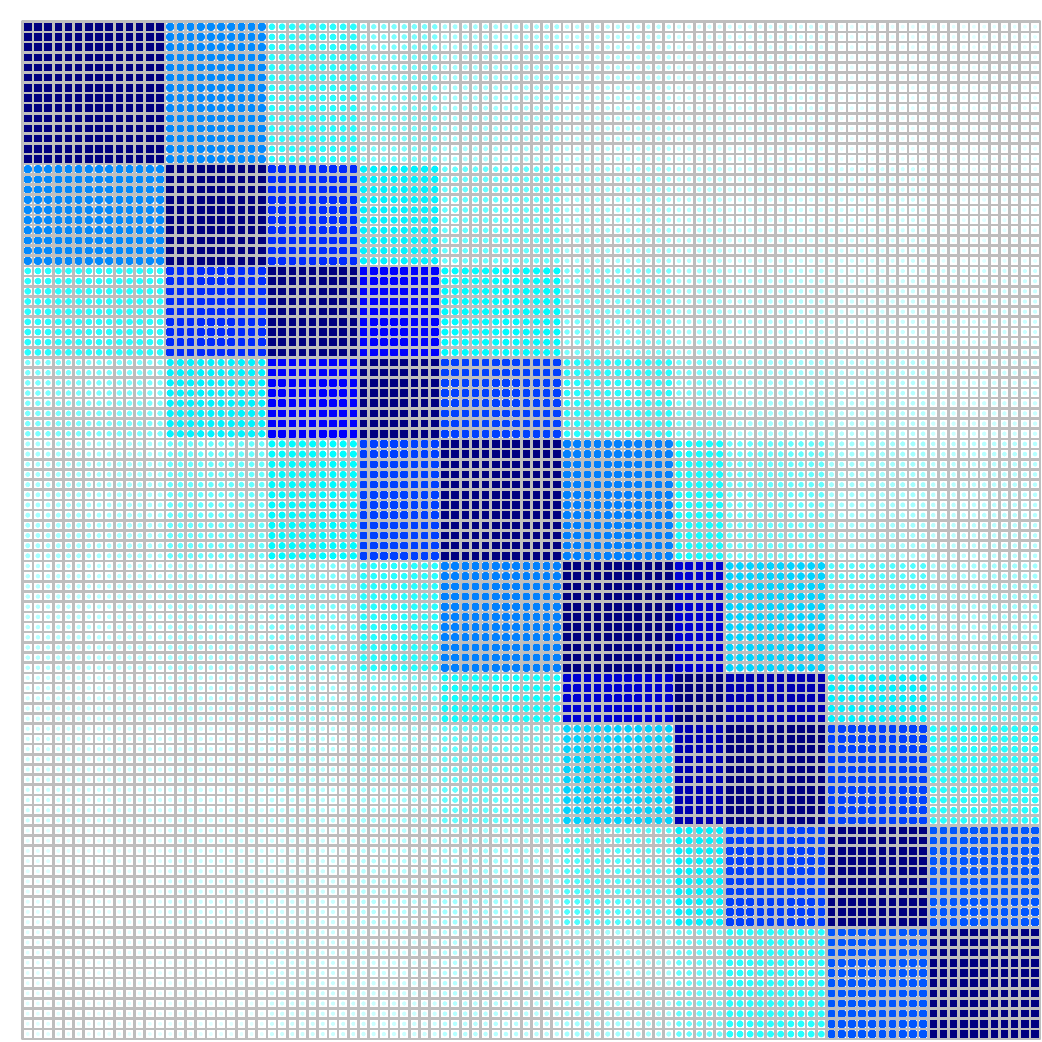}}
		\caption{The left panel gives an example of the banding structure of $\Omega^s$ suggested in (\ref{eq: Omega banding structure}) and the right panel correspond to the structure of $\Wscr^s$ given in Lemma \ref{lemma: F to H}.}
		\label{fig: F to H}
	\end{center}
\end{figure}

	We further leverage the prior knowledge on the distance between nodes under the assumption that nodes further apart are less likely to be grouped together. Specifically, let $d: \Vcal \times \Vcal \mapsto [0,\infty)$ be the distance metric, which satisfies the well-known non-negativity, symmetry and identity property: $\forall (v_i,v_j) \in \Vcal \times \Vcal$,
	\begin{eq*}
		d_{ij} \equiv d(v_i,v_j) \geq 0,\quad d(v_i,v_j) = d(v_j,v_i), \quad d(v_i,v_j) =0 \Leftrightarrow v_i = v_j
	\end{eq*}
    Here the triangular inequality assumption on $d(\cdot,\cdot)$ is not required.
	For example, if $\Vcal$ denotes all strings of ICD9 codes, one simple way to define the distance metric is based on their numerical representations, e.g., $d(``331.1", ``331.9")=|331.1 - 331.9|=0.8$.  An alternative distance is based on the number of steps needed to connect two codes on the ICD9 hierarchy tree.  For $k=1, ..., K$, we define the centroid node of the cluster $\Vcal_k$ as $v_{c_k}$, where 
	\begin{eq*}
	c_k := \min\left\{i: \argmin_{v_i \in \Vcal_k} \sum_{v_j \in \Vcal_k} d_{ij} \right\}  
	\end{eq*}
	and $n_k := |\Vcal_k|$ is the cardinality of  $\Vcal_k$. 
	To leverage the prior knowledge that nodes further apart less likely belong to the same group, we assume that 
	\begin{itemize}		
		\item[] ({\sf C1}) \em There exists some $\delta>0$ such that $d(v_i, v_{c_k}) \leq \delta,\quad \forall v_i \in \Vcal_k,\quad  k=1,\ldots, K$.
	\end{itemize}
	Obviously, $\delta$ is the radius of each cluster.  
	More than that, it can be used as a lever to characterize the confidence level cast on the prior distance knowledge. More precisely, a smaller $\delta$ leads to more engagement of the prior knowledge while a larger $\delta$ downplays its role. More discussions on the effect of $\delta$ in our methodolgy will be given in Section 3. The adoption of ({\sf C1}) enables us to employ thresholding on any pairwise similarity whose pairwise distance is beyond $2\delta$.  For example, in Figure \ref{fig: delta},  $\delta_2$ only keeps within-in cluster pairwise similarities while $\delta_1$ entails no thresholding since all nodes are encompassed in the outter black dashed circle. 
	Without loss of generality, we assume $\delta > d_0:= \min_{1\leq i < j \leq n} d_{ij}$  which is naturally satisfied provided $K < n$. Furthermore, we assume that the magnitude of $\Omega_{kl}^s$ shall vanish as $v_{c_{k}}$ and $v_{c_{l}}$ become more and more distant and hence consider each $\Omega^s$ reside in the following restricted class of matrices, inspired by \cite{bickel2008regularized},
	\begin{align}
	\label{eq: Omega banding structure}
\Fcal_{\alpha_s} := \Fcal(\alpha_s, L, d_0,\beta) = \biggl\{ &   \Obb =[O_{kl}]_{k=1,...,K}^{l = 1,...,K} \in [-L,L]^{K\times K}: \ \max_{1\leq k \leq K} \sum_{l: d(v_{c_k},v_{c_l}) > h}  |O_{kl}|  \leq L(h/d_0)^{-\alpha_s} ,  \nonumber \\ 
& \hspace{.4in} \ \Obb=\Obb\trans, \;  \mbox{for all }  h > 0,  \  0 < \beta \leq \gamma_{K}(\Obb) \leq  \gamma_1(\Obb) \leq \beta^{-1}  \biggl. \biggr \}	
	\end{align} 
	where $\gamma_{k}(\Obb)$is $k$-th largest absolute eigenvalue of $\Obb$, and $\alpha_s \geq 0$ controls the vanishing rate for off-diagonal entries as they move away from the diagonals guided by the distance. Intuitively, the larger $\alpha_s$ is, the easier to distinguish different communities. It is easy to see that $\alpha_1 < \alpha_2$ implies that $\Fcal_{\alpha_2} \subset \Fcal_{\alpha_1}$, and $\alpha=0$ reduces to the null case where the decaying phenomenon is abscent. Restriction of $\Omega^s$ on $\Fcal_{\alpha_s}$ leads to an immediate consequence on the parameter space of $\Wscr^s$ described in Lemma \ref{lemma: F to H}, which loosely speaking is an expanded copy of $\Fcal_{\alpha_s},$ by the largest cluster size. As a special case, Figure \ref{fig: F to H} gives an illustration on the structure of $\Omega^s$ and its effect on $\Wscr^s$ when nodes happen to follow a natural ordering. 
\begin{lemma}
	\label{lemma: F to H}
	Suppose $\Omega^s \in \Fcal_{\alpha_s}$ defined in (\ref{eq: Omega banding structure}) and $\Vcal$ satisfies condition ({\sf C1}), given the membership matrix $\Zbb^*$, then $\Wscr^s:= \Zbb^*\Omega^s\Zbb^{*\trans}$ resides in the following class of matrices
	\begin{eq}
		\label{eq: W banding structure}
		\Hcal_{\alpha_s} := \Hcal(\alpha_s,\delta,L,d_0,\beta) = \left\{\Abb =[A_{ij}]_{i=1,...,n}^{j=1,...,n} \in [-L,L]^{n \times n}: \Abb=\Abb\trans,\; \max_{1\leq i \leq n} \sum_{j:d(v_i, v_j) > h}  |A_{ij}| \leq \right. \\ L n_{max}\left(\frac{h-2\delta}{d_0} \right)^{-\alpha_s}, \mbox{ for all } h > 2\delta,  \mbox{ and }  0 < n_{min}\beta \leq \gamma_{K}(\Abb) \leq  \gamma_1(\Abb) \leq n_{max}/\beta  \biggl. \biggr \}	
	\end{eq} where $n_{min}=\min_k n_k, n_{max}=\max_k n_k$.
\end{lemma}

	\section{Multi-view banded spectral clustering (mvBSC)}    
	In this section, we first summarize our proposed mvBSC procedure in Algorithm \ref{alg: mvclust banding} and then give a brief explanation on the reasoning behind it. We then discuss the theoretical results supporting the validity and optimality of the proposed algorithm as well as the appropriate choices of the banding vector $\bh=(h_1,...,h_m)\trans$ and the weighting vector $\blambda = (\lambda_1, ..., \lambda_m)\trans$. Without loss of generality, we restrict our attention to the convex combination, $\blambda \geq 0, \one_m\trans \blambda = 1$.

	\subsection{The mvBSC procedure}
	The proposed mvBSC procedure is summarized as follows:
	
\begin{algorithm}[H]
	\caption{Multi-view banded spectral clustering (mvBSC)}
	\label{alg: mvclust banding}
	
	\KwIn{$m$ $n \times n$ similarity matrices $\Wbb^1, \ldots, \Wbb^m$, pairwise distances $\{d_{ij}\}_{i=1,...,n}^{j=1,...,n}$, number of groups $K$, a set of banding parameters $h_1, ..., h_m$, and a set of weighting parameters $\lambda_1,\ldots, \lambda_m \geq 0, \sum_{s=1}^m \lambda_s = 1$.}
	
	\KwOut{Membership matrix $\Zbbhat^*_{\blambda} \in \Zscr_{n,K}$}
	
	\For{s = 1:m}{
		
		(1) $B_{h_s}(\Wbb^s):= [\Wbb_{ij}^s I(d_{ij} \leq h_s)]_{i=1,...,n}^{j=1,...,n} \leftarrow$ banding $\Wbb^s$ using the banding parameter $h_s$.	
		
		(2) $\Ubbhat^s \leftarrow $ the matrix of concatenating $K$ eigenvectors corresponding to the first $K$ largest absolute eigenvalues of $B_{h_s}(\Wbb^s)$.
		
	}
	$\Ubbhat_{\blambda}^* \leftarrow$ the matrix of concatenating $K$ eigenvectors corresponding to the first $K$ largest eigenvalues of  $\sum_{s=1}^m \lambda_s \Ubbhat^s \Ubbhat^{s\trans}$.
	
	Treating each row of $\Ubbhat_{\blambda}^*$ as a point in $\Rcal^K$, run k-means to cluster these $n$ points into $K$ groups and obtain a corresponding membership matrix $\Zbbhat^*_{\blambda}$.
	
	\KwResult{$\Zbbhat^*_{\blambda}$}
	
\end{algorithm}

\vspace*{3mm}

	Algorithm \ref{alg: mvclust banding} involves two key steps: operating banding to each similarity matrix $\Wbb^s$ and performing spectral clustering on a convex combination of $m$ projector matrices $\Ubbhat^s \Ubbhat^{s\trans}$. To see the reasoning, we first examine the eigen-space of $\Wscr^s$.  Let 
	$$\Delta = \left[\diag(\one_n \trans \Zbb^*)\right]^{1/2} = \diag(\sqrt{n_1}, \ldots,  \sqrt{n_K}), \quad \Ubb^* = \Zbb^*\Delta^{-1},
	$$ 
and $\Omegatilde^s := \Delta \Omega^s \Delta = [\Omega_{kl}\sqrt{n_kn_l}]_{k=1,\cdots, m}^{l=1,\cdots, m}$ with eigen-decomposition  $\Omegatilde^s = \Vbb^s \Dbb^s\Vbb^{s\trans}$. Then the eigen-decomposition of $\Wscr^s$ is given by 
$$\Wscr^s  = \Ubb^s \Dbb^s \Ubb^{s\trans} = \Ubb^*(\Vbb^s\Dbb^s\Vbb^{s\trans})\Ubb^{*\trans}, \quad \mbox{where}\ \Ubb^s = \Zbb^*\Delta^{-1}\Vbb^s=\Ubb^*\Vbb^s.$$ 
Clearly, $\Ubb^s$ is a rotation of $\Ubb^*$ by $\Vbb^s$ and they correspond to the same $K$-dimensional subspace expressed by $P_{\Zbb^*} = \Ubb^*\Ubb^{*\trans} = \Ubb^s\Ubb^{s\trans}$.  Since $(\Delta^{-1}\Vbb^s)^{-1}$ exists, $\Ubb^s_{i\cdot} = \Ubb^s_{j\cdot} \Leftrightarrow g_i = g_j$, meaning that $v_i$ and $v_j$ are in the same group if and only if their corresponding rows in $\Ubb^s$ are the same. So the eigen-space of $\Wscr^s$ is membership structured and recovering $\Ubb^s$ is equivalent as recovering $\Zbb^*$.  

Now consider the difference between $\Wbb^s$ and $\Wscr^s$: 
	$$\Wbb^s - \Wscr^s = (\Wbb^s - \sE_{\Zbb^*} \Wbb^s) + \diag(\sE_{\Zbb^*} \Wbb^s-\Wscr^s) = (\Wbb^s - \sE_{\Zbb^*} \Wbb^s) + \diag(\Wbb^s-\Wscr^s), $$ which is a symmetric bounded noise matrix plus a diagonal matrix. For the above equality, we note that $\diag(\Wbb^s) = \diag(\sE_{\Zbb^*}\Wbb)$ because the diagonals of $\Wbb^s$ are constants. Although the deviation from the eigenspace of $\Wbb^s$ to that of $\Wscr^s$ is always upper bounded by the operator norm of their difference, with no further information on the structure of $\Wscr^s$, one can at most know the operator norm deviation is on the scale of $\sqrt{n }$ since the noise matrix $(\Wbb^s - \sE_{\Zbb^*} \Wbb^s)$ is on this scale and the remaining diagonal matrix can be treated as a constant and hence makes negligible contributions. Fortunately, Theorem \ref{thm: optimal bandwidth}  sheds light on the benefit of banding to significantly reduce the upper bound to the scale of $ \max \left\{\left( n_{max}^{{1 \over 2\alpha_s + 1}}(\log n)^{{\alpha_s \over 2\alpha_s + 1}} \right), \sqrt{{\delta \over d_0} \log n },\log n \right\}$.  Therefore, the space spanned by $\Ubbhat^s$ would show more resemblance to the space spanned by $\Ubb^*$.

	The other key step is the use of a weighted average $\sum_{s=1}^m \lambda_s \Ubbhat^s \Ubbhat^{s\trans}$ to estimate $P_{\Zbb^*}$.   Since each $\Ubbhat^s \Ubbhat^{s\trans}$ can be viewed as a stand-alone estimator of $P_{\Zbb^*}$, $\Ubbhat^s \Ubbhat^{s\trans}$ can be decomposed into $P_{\Zbb^*}+ \Ebb^s$, where $\Ebb^s$ is a symmetric error matrix. As an analogy of a weighted least square problem, to allow for heterogeity in noise corruptions, given a weight vector $\blambda$, it is ideal to find
	\begin{eq}
		\label{eq: weighted obj}
		\widetilde{\Zbb}^*_{\blambda} = \argmin_{\Zbb \in \Zscr_{n,K}} \sum_{s=1}^m \lambda_s \| \Ubbhat^s \Ubbhat^{s\trans} - P_{\Zbb} \|^2_F
	\end{eq}
	From a regularization perspective, these weights essentially put penalty on each view so that they can be dragged to a common ground to maximize consensus expressed by $\widetilde{\Zbb}^*_{\blambda}$. However, the solution of (\ref{eq: weighted obj}) is NP-hard to find, and the alternative is to find the un-constrained solution $\sum_{s=1}^m \lambda_s \Ubbhat^s \Ubbhat^{s\trans}$ as a surrogate whose eigenvectors can be subsequently used to reconstruct $\Zbb^*$. 
	
	In summary, the two sets of parameters $\bh$ and $\blambda$ in the proposed mvBSC procedure play orthogonal but complementary roles. The banding parameter $h_s$ maximally attempts to improve each individual performance, while the weight parameter $\lambda_s$ allows for efficient messsage passing horizontally to make up each other's deficiencies.

	\subsection{Error analysis}
	To study the statistical performance of the proposed mvBSC procedure(Algorithm \ref{alg: mvclust banding}), it is important to realize that errors consist of two parts--distance from $\Ubb^*$ to $\Ubbhat_{\blambda}^*$ and the membership misallocation arising from the k-means step. The minimal distance between $\Ubb^*$ and $\Ubbhat^*_{\blambda}$ is equivalent to the distance between their respective subspace $P_{\Zbb^*}$ and $\Ubbhat^*_{\blambda}\Ubbhat^{*\trans}_{\blambda}$ \citep{vu2013minimax} 
	\begin{equation}
		\label{eq: subspace dist}
		{1 \over 2} \inf_{Q} \|\Ubbhat^*_{\blambda} - \Ubb^*Q\|^2_F \leq \| \Ubbhat^*_{\blambda}\Ubbhat^{*\trans}_{\blambda} - P_{\Zbb^*}\|^2_F \leq \inf_{Q}\|\Ubbhat^*_{\blambda} - \Ubb^*Q\|^2_F
	\end{equation}
 where $Q$ is a $K \times K$ orthonormal matrix.  
 To evaluate the quality of the final k-means step, it is necessary to first define a ``mis-clustered" node. To this end,  recall that the k-means obtains 
	\begin{equation}
		\label{eq: kmeans opt kit}
		(\Zbbhat^*_{\blambda}, \Abbhat_{\blambda} ) = \argmin_{\Zbb \in \Zscr_{n,K}, \Abb \in \Rcal^{K \times K}} \norm{\Ubbhat^*_{\blambda} - \Zbb \Abb}_F^2
	\end{equation}
	in which $\Abbhat_{\blambda_{k\cdot}}\trans \in \Rcal^K$ represents the $k$-th centroid \citep{steinhaus1956division}. Intuitively,  if $\Ubb^*_{i\cdot}Q$ is closer to $\Abbhat_{\blambda_{g_i \cdot}}$ than it is to any other $\Abbhat_{\blambda_{k\cdot}}$ for $k \neq g_i$, then node $v_i$ can be correctly clustered.   
	
\begin{definition}[set of mis-clustered nodes]
	\label{def: mis-clustered nodes}
	Given $\Zbb^* \in \Zscr_{n,K}$, let $\Ubb^* = \Zbb^*\Delta^{-1}$, $\Ubbhat_{\blambda}^*$ given in Algorithm \ref{alg: mvclust banding} , $\Abbhat_{\blambda}$ defined in (\ref{eq: kmeans opt kit}) and $Q$ be a $K \times K$ orthonormal matrix satisfying (\ref{eq: subspace dist}), define $\Mcal_{\blambda}$ as the set of mis-clustered nodes
	$$\Mcal_{\blambda} =\biggl \{ i : \|\Abbhat_{\blambda_{g_i\cdot}} - \Ubb^*_{i\cdot} Q \|_2 \geq  \sqrt{1/2n_{max}} \biggr\}$$
\end{definition}
\begin{rmk}
	The definition of $\Mcal_{\blambda}$ is a sufficient condition to ensure $\| \Abbhat_{\blambda_{g_i \cdot}} -  \Ubb^*_{i \cdot} Q\|_2 \leq \|\Abbhat_{\blambda_{g_i \cdot}} - \Ubb^*_{j\cdot} Q\|_2$ for any $g_j \neq g_i$, which was firstly considered in \cite{rohe2011spectral}. 	The error analysis in this paper mainly addresses the global optimum of (\ref{eq: kmeans opt kit}), and this optimization problem could suffer from local optima in practice.
\end{rmk}

\begin{theorem}[Optimal Choice of Banding Parameter $h$]
	\label{thm: optimal bandwidth}
	
	Given the membership matrix $\Zbb^* \in \Zscr_{n,K}$, consider a similarity matrix $\Wbb \in [-L,L]^{n \times n}$, generated according to (\ref{eq: model}), in which $\Wscr = \Zbb^* \Omega \Zbb^{*\trans}$ and $\Omega \in \Fcal_{\alpha}$ defined in (\ref{eq: Omega banding structure}), if ${h - 2\delta \over d_0} \asymp  \left( \frac{n_{max}}{\sqrt{\log n} } \right)^{\frac{2}{ 2\alpha + 1}} $, the mean absolute operator-norm error loss
	
	$$\sE_{\Zbb^*} \norm{ B_{h}(\Wbb) - \Wscr}  \lesssim \max \left\{\left( n_{max}^{{1 \over 2\alpha + 1}}(\log n)^{{\alpha \over 2\alpha + 1}} \right), \sqrt{{\delta \over d_0} \log n },\log n \right\} $$ 
	More specifically, if $n_{max} \gtrsim (\log n)^{\alpha+1}$,
		$$\sE_{\Zbb^*} \norm{ B_{h_s}(\Wbb) - \Wscr }  \lesssim \left\{\begin{array}{cc} \left( n_{max}^{{1 \over 2\alpha + 1}}(\log n)^{{\alpha \over 2\alpha + 1}} \right) & \mbox{if } \; {\delta \over d_0} = o\left(  \left( \frac{n_{max}}{\sqrt{\log n} } \right)^{\frac{2}{ 2\alpha + 1}} \right) \\	
	\sqrt{{\delta \over d_0} \log n } & \mbox{otherwise} 
	\end{array}  \right. $$
	and for a finely-sliced network, i.e. $n_{max} = o\left( [\log n]^{\alpha+1}\right)$,
		$$\sE_{\Zbb^*} \norm{ B_{h}(\Wbb) - \Wscr }  \lesssim \left\{\begin{array}{cc} \sqrt{{\delta \over d_0} \log n } & \mbox{if } \; {\delta \over d_0} \gtrsim  \log n  \\	
 \log n & \mbox{otherwise} 
\end{array}  \right. $$	

\end{theorem}

 Theorem \ref{thm: optimal bandwidth} explicitly shows the effect of banding in reducing the mean absolute operator-norm error loss whose upper bound is a function of the key model parameters $(n, n_{max}, \delta, \alpha)$. The optimal banding parameter $h$ is reflective of the decay rate in the sense that a smaller $\alpha$ yields a larger $h$. Adopting a conservative $\delta$ would lead the mean absolute operator-norm error loss to the order of $\sqrt{ {\delta \over d_0}\log n}$. In the extreme scenario where $\frac{\delta}{d_0} \lesssim n$, suggesting that one is too conservative to perform banding, the mean absolute operator-norm error loss given in Theorem \ref{thm: optimal bandwidth} is upper bounded by $\sqrt{n\log n}$, reducing to the standard result up to a $\sqrt{\log n}$ factor on a random matrix whose entries are independent copies of a random variable with zero mean, unit variance, and finite fourth moment.

\begin{theorem}[mis-clustered error rate]
	\label{thm: mis-clustered error rate}
	Given the membership matrix $\Zbb^* \in \Zscr_{n,K}$, consider a sequence of similarity matrices $\Wbb^1, \ldots, \Wbb^m$, generated independently according to (\ref{eq: model}), in which $\Wscr^s = \Zbb^* \Omega^s \Zbb^{*\trans}$ and $\Omega^s \in \Fcal_{\alpha_s}$ defined in (\ref{eq: Omega banding structure}), and $\gamma^s_{n,K} = \gamma_K(\Wscr^s)$. Given a weighting vector $\blambda$, let $\Mcal_{\blambda}$ be the set of mis-clustered nodes given in Definition \ref{def: mis-clustered nodes}, then with probability at least $1-m/n$,
	$$ 	\frac{|\Mcal_{\blambda}|}{n} = O_p\left[ {n_{max} \over n} \sum_{s=1}^m \left( \frac{\lambda_s}{\gamma^{s}_{n,K}}\right)^2 \max \left\{ n_{max}^{{2 \over 2\alpha_s + 1}}(\log n)^{{2\alpha_s \over 2\alpha_s + 1}} , \frac{\delta}{d_0}\log n,(\log n)^2 \right\} \right]$$

\end{theorem}

\begin{corollary}
	\label{cor: special case}
	Suppose $\Omega_1, ..., \Omega_m \in \Fcal_{\alpha}$ for some $\alpha \geq 0$, and the underlying $\Zbb^*$ exhibits a balanced community structure, i.e., $ n_{max} \asymp n_{min}$, then with probability at least $1-m/n$,
	\begin{itemize}
		\item[(i)] if $n_{max} \gtrsim (\log n)^{\alpha+1}$, 	
		$$\frac{|\Mcal_{\blambda}|}{n} = O_p \left [ \frac{\log n}{n n_{max}}  \max\left\{ \left( \frac{n_{max}}{\sqrt{\log n}}\right) ^{{2 \over 2\alpha+1}}, {\delta \over d_0}\right\} \right]$$
		
		\item[(ii)] if $n_{max} = o\left( [\log n]^{\alpha+1}\right)$, 
		$$ \frac{|\Mcal_{\blambda}|}{n} = O_p \left\{ \frac{\log n}{n n_{max} } \max\left( {\delta \over d_0}, \log n\right)\right\} $$
	\end{itemize}
	
\end{corollary}

\subsection{Optimal choice of $\{\lambda_s\}_{s=1}^m$}
	
A remaining important question is that in what sense these $m$ views can be optimally combined via the weighting parameters $\blambda=(\lambda_1, ..., \lambda_m)\trans$? Ideally, a desirable set of weighting parameters $\blambda^*$ 
shall minimize the population-level mis-clustered node size $\sE_{\Zbb^*} |\Mcal_{\blambda}|$  and thus can be referred as the oracle weighting vector.
Despite its attractiveness, an explicit form of $\sE_{\Zbb^*} |\Mcal_{\blambda}|$ is intractable due to the difficulty in deriving the deviation from $\Ubbhat^s \Ubbhat^{s \trans}$ to  $P_{\Zbb^*}$. As an alternative strategy, we seek to derive an upper bound of $\sE_{\Zbb^*}|\Mcal_{\blambda}|$ as a surrogate objective function, $q_{\bh}(\blambda)$, that leads to an approximately optimal solution that sufficiently reflects all $m$ views. 
To this end, we note that from the proof of Theorem \ref{thm: mis-clustered error rate},  $|\Mcal_{\blambda}|$ is upper-bounded by $\| \sum_{s=1}^m \lambda_s \Ubbhat^s \Ubbhat^{s\trans}- P_{\Zbb^*} \|^2$ up to a constant, and 
\begin{equation}
	\label{eq: lei}	
	\norm{ \sum_{s=1}^m \lambda_s \Ubbhat^s \Ubbhat^{s\trans}- P_{\Zbb^*} }^2 \leq  m  \sum_{s=1}^m \lambda^2_s \|\Ubbhat^s\Ubbhat^{s\trans} - \Ubb^s \Ubb^{s \trans}\|^2  \leq 4m \sum_{s=1}^m \left(\frac{\lambda_s}{\gamma^s_n}\right)^2 \|B_{h_s}(\Wbb^s) - \Wscr^s\|^2 
\end{equation}
It thus suffices to derive an upper bound for each individual $\sE_{\Zbb^*} \norm{ B_{h_s}(\Wbb^s) - \Wscr^s }^2 $ as given below.

\begin{theorem}
	\label{thm: q_lambda}
	Suppose $n_{max} \gtrsim (\log n)^{\alpha_s+1}$, and choose $h_s = 2\delta + d_0\left( \frac{Ln_{max} }{2\sqrt{\log n}}\right)^{\frac{2}{2\alpha_s + 1}}, \; s=1,...,m$, then for some constant $C>0$,
	$$\sE_{\Zbb^*} \norm{ B_{h_s}(\Wbb^s) - \Wscr^s }^2 \leq C h_s \sigma_s^2 \log n$$
	and 	
\begin{eq*}
	\sum_{s=1}^m \left(\frac{\lambda_s}{\gamma^s_{n,K}}\right)^2 \|B_{h_s}(\Wbb^s) - \Wscr^s\|^2 &\leq C \log n \sum_{s=1}^m \left(\frac{\lambda_s \sigma_s}{\gamma^s_{n,K}}\right)^2  h_s 
\end{eq*} 
\end{theorem}

Theorem \ref{thm: q_lambda} immediately implies that the surrogate objective function can be given as
\begin{equation}
	\label{eq: q_obj}
	q_{\bh}(\blambda) = \sum_{s=1}^m \left(\frac{\lambda_s \sigma_s}{\gamma^s_{n,K}}\right)^2   h_s
\end{equation}
It is straightforward to see that $\blambda_q^* = \argmin_{\blambda: \sum_{s=1}^m \lambda_s =1, \lambda_s \geq 0}q_{\bh}(\blambda) $ takes the form
\begin{eq}
	\label{eq: best.q}
\lambda_{qs}^* = h_s^{-1}\left( \frac{\gamma^s_{n,K}}{\sigma_s}\right)^2  \left [\sum_{t=1}^m h_t^{-1}\left(\frac{\gamma_{n,K}^t}{\sigma_t}\right)^2\right ] ^{-1},  \; s =1,...,m. 
\end{eq}
Furthermore, if $\Omega_1,..., \Omega_m \in \Fcal_{\alpha}$ for some common $\alpha >0$, 
\begin{eq}
	\label{eq: snr}
\lambda_{qs}^* = \left( \frac{\gamma^s_{n,K}}{\sigma_s}\right)^2  \left[  \sum_{t=1}^m \left(\frac{\gamma_{n,K}^t}{\sigma_t}\right)^2\right]^{-1},  \; s =1,...,m.
\end{eq}
It is easy to see from (\ref{eq: snr}), $\lambda^*_{qs} \propto (\gamma_{n,K}^{s}/\sigma_s)^2$, which can be comprehended as a measure of signal-to-noise ratio (SNR) since $\gamma_{n,K}^s$ characterizes the capability of $\Wscr^s$ unveiling $\Zbb^*$ at the population level whereas $\sigma_s$ summarizes the sample-level ($\Wbb^s$) corruption. This SNR flavored weighting scheme is seamlessly aligned with one's intuition that quality evaluation on each view shall  consider its inherent ability and external noise extent simultaneously. To use it in practice, $\gamma_{n,K}^s$ can be estimated by the largest $K$-th eigenvalue of $B_{h_s}(\Wbb^s)$ and $\sigma_s$ can be estimated as in Remark \ref{rmk: sigma estimation}. More generally, (\ref{eq: best.q}) takes $\alpha_s$ into account through $h_s^{-1}$, which is also intuitive in the sense that it downweights the view with a slower decay rate. Although (\ref{eq: snr}) is only a special case of (\ref{eq: best.q}), it is often the case in practice that $\alpha_s$'s are very close or researchers prefer to take a bit more conservative perspective($\alpha_1 \leq \alpha_2$ implies that $\Fcal_{\alpha_2} \subseteq \Fcal_{\alpha_1}$), in whichever case $h_s$ makes negligible difference and (\ref{eq: best.q}) reduces to (\ref{eq: snr}). In the sequel, we refer mvBSC$\subSNR$ to the mvBSC using (\ref{eq: snr}) and mvBSC$_q$ to the mvBSC using (\ref{eq: best.q}). 

\begin{rmk}[Estimation of $\sigma_s^2$]	
	\label{rmk: sigma estimation}
Recall that the variation in each similarity matrix consists of two sources: within group variation and across group variation. Let $\MS^2_{within}$ and  $\MS^2_{across}$ denote the mean square error respectively.  Given a membership matrix $\Zbb \in \Zscr_{n,K}$,
\begin{eq}
\MS^2_{within} &= \sum_{k=1}^K {1 \over n_k(n_k-1)/2 -1} \sum_{\{(i,j): g_i =g_j=k, i<j\}} (W^s_{ij} - \widehat{\Omega}^s_{kk})^2 \\
\MS^2_{across} &= \sum_{1\leq k < l \leq K}   {1\over n_kn_l -1}  \sum_{\{(i,j): g_i=k, g_j=l\}} (W^s_{ij} - \widehat{\Omega}^s_{kl})^2
\end{eq} where $\widehat{\Omega}^s_{kk} =  {2 \over n_k(n_k-1)} \sum_{\{(i,j): g_i =g_j=k, i<j\}} W^s_{ij}$ and $\widehat{\Omega}^s_{kl} = {1\over n_kn_l}  \sum_{\{(i,j): g_i=k, g_j=l\}} W^s_{ij}$.

Thus, $\sigma_s^2$ can be estimated by 
\begin{eq}
\label{eq: sigma hat}
\hat{\sigma}_s^2 = {2 \over K(K+1)} \biggl (\MS^2_{within} + \MS^2_{across}\biggr )
\end{eq}
\end{rmk}

\def\Xcal{\mathcal{X}}
\def\Ycal{\mathcal{Y}}

\section{Simulations}
\begin{figure}[htbp]
	\begin{center}
		\includegraphics[width=7cm,height=6cm]{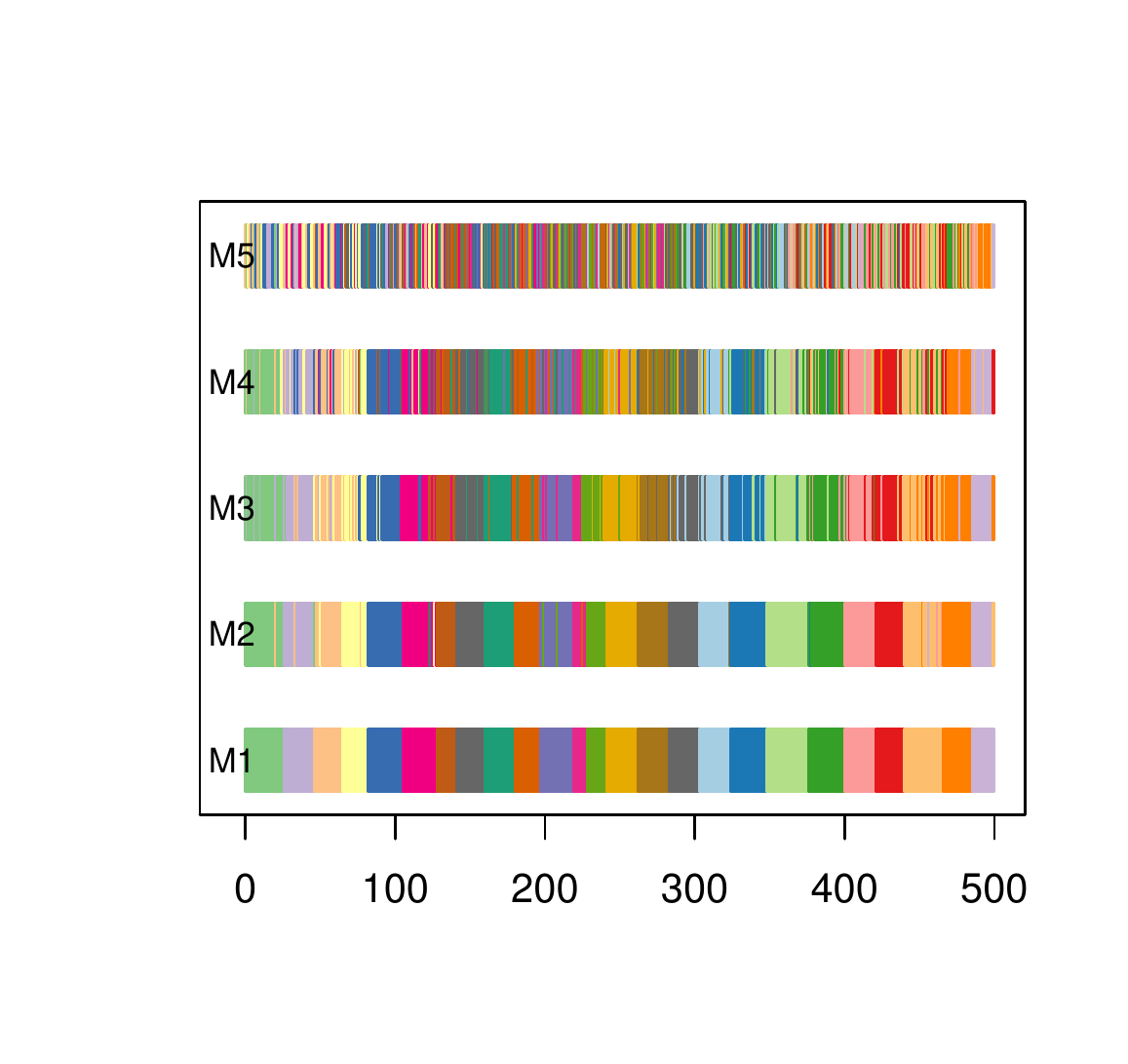}
		
		\caption{Graphical representation of the membership matrix $\Zbb^*$ under (M1), (M2), (M3), (M4) and (M5) with groups indexed by colors.}
		\label{fig: models}		
	\end{center}
\end{figure}

We have performed extensive simulation studies to examine the finite sample clustering performance of the proposed mvBSC method.  Throughout simulations, we let $m=2$, $d(v_i,v_j)=|i-j|/10$ and use $v_i$ and its index $i$ exchangeably. To examine the robustness of mvBSC to the underlying network structure,  we considered five different models (M1)-(M5) for the membership matrix $\Zbb^*$ as illustrated in Figure  \ref{fig: models} where nodes from the same cluster share the same color. In (M1), the clusters have a clear block structure with a total of 25 clusters and cluster size ranging from 9 to 28. Under (M1), our model assumption (C1) holds exactly with a small $\delta$.  We then gradually depart from this assumption by perturbing (M1). More specifically, for $k=2,3,4,5$, $\Zbb^*$ in model (Mk) is generated by randomly swapping the node's membership in (M1) to one of the $l_k$ most adjacent clusters with probability $p_k$, where we let $(p_2, l_2) = (0.01, 4)$, $(p_3,l_3)=(0.1, 2)$, $(p_4, l_4) = (0.05, 6)$ and $(p_5, l_5) = (0.1, 8)$. 
Thus $p_k$ and $l_k$ jointly control the degree of departure from assumption (C1) and the assumption no longer holds for a finite constant $\delta$ in the most challenging case of (M5). 
Throughout, we let $n = 500$ and for each membership model, we considered $K=50, 25, 10$ to reflect a small, medium, large average cluster size.  

For a given membership matrix $\Zbb^*$ and its associated network partition $\Vcal = \cup_{k=1}^K \Vcal_k$ , we let $c_k = {1 \over |\Vcal_k|} \sum_{i: \Zbb^*_{ik}=1} i$ denote the centroid node of $\Vcal_k$. 
 We then generate each population level similarity matrix $\Wscr^s$ independently according to (\ref{eq: model}) from $\Omega^s \in \Fcal_{\alpha_s}$ with $\alpha_1 = 0.4$ and $\alpha_2 = 0.6$, where
\begin{equation}
\Omega^s_{kl} = \left \{ \begin{array}{lcc}
1 & \mbox{ if }  & k =l \\
0.6 |c_k-c_l|^{-(\alpha_s+1)}  & \mbox{ if } & k \neq l \end{array}
\right .  
\end{equation}  
The observed similarity matrix $\Wbb^s$ was generated as $W_{ij}^s = \{(\Wscr_{ij}^s + \epsilon_{ij}^s)\wedge 1\}\vee(-1)$, where $\epsilon_{ij}^s \sim N(0, \sigma_s^2)$. We considered three noise levels with $(\sigma_1, \sigma_2)$ being (i) $(0.2,0.4)$ for low noise; (ii) $(0.4,0.6)$ for median noise; and (iii) $(0.6,0.8)$ for high noise.  Here, $W_{ij}^s$ is banded to be between $[-1,1]$ to mimic the cosine similarity used in the real data example and we let all diagonals of $\Wbb^s$ to be 1. We choose $h_s = 2\delta + d_0\left( \frac{Ln_{max}}{\sqrt{\log n}}\right)^{\frac{2}{2\alpha_s + 1}}, s=1,2$ suggested in Theorem \ref{thm: q_lambda}. 

For each configuration, we performed clustering based on our proposed mvBSC method as well as a few existing methods including spectral clustering using (i) kernel addition (KA) matrix $\Wbb^+ = \Wbb^1 + \Wbb^2$ (ii) the Laplacian of $\Wbb^+$ (KAL); (iii) the normalized Laplacian of $\Wbb^+$ (normKAL); (iv) each single $\Wbb^s$ alone (singleW);  (v) the Laplacian of each single $\Wbb^s$ (singleL); and (vi) the normalized Laplacian of each single $\Wbb^s$. Here, for a $\Wbb$ under consideration, the Laplacian matrix is derived using $\Wbb - \min(\Wbb)$ since it is defined based on a non-negative adjacency matrix. For the mvBSC method, we considered different approaches to select $\blambda$ including mvBSC$_q$ and mvBSC$\subSNR$ as well as an oracle method that chooses $\blambda$ by minimizing the mis-clustering rate. For each configuration and each clustering method, we quantify the quality of the clustering based on the average clustering accuracy, defined as one minus the mis-clustered error rate, and the normalized mutual information (NMI) over 100 replications. The NMI is a commonly used measure in the networks literature and is known to be impartial with respect to $K$ \citep{strehl2002cluster}. Specifically, 
given a vertex set $\Vcal=\{v_i\}_{i=1}^n$, the NMI between a partition $\Xcal$ with $\Vcal= \Vcal_1^{\Xcal} \cup \cdots \cup \Vcal_{K_\Xcal}^{\Xcal}$ and a gold standard reference partition $\Xcal_0$ 
with $\Vcal= \Vcal_1^{\Xcal_0} \cup \cdots \cup \Vcal_{K_{\Xcal_0}}^{\Xcal_0}$ is
\begin{equation}
\label{eq: NMI}
\NMI(\Xcal, \Xcal_0) = \frac{\sum_{k=1}^{K_{\Xcal}}\sum_{l=1}^{K_{\Xcal_0}} \left|\Vcal_k^{\Xcal} \cap \Vcal_l^{\Xcal_0}\right| \log \left(\frac{\left|\Vcal_k^{\Xcal} \cap \Vcal_l^{\Xcal_0}\right|}{\left|\Vcal_k^{\Xcal}\right|\left|\Vcal_l^{\Xcal_0}\right|}\right)}{\sqrt{\left\{\sum_{k=1}^{K_{\Xcal}} \left|\Vcal_k^{\Xcal}\right| \log \left(\frac{\left|\Vcal_k^{\Xcal}\right|}{n}\right)\right\}\left\{\sum_{l=1}^{K_{\Xcal_0}} \left|\Vcal_l^{\Xcal_0}\right|\log \left(\frac{\left|\Vcal_l^{\Xcal_0}\right|}{n}\right)\right\} }} ,
\end{equation}
which is a score ranging from 0 to 1 with a higher value indicating that $\Xcal$ is more similar to the reference partition $\Xcal_0$. We let $\Xcal_0$ be the true underlying partition in our simulation studies and suppress the dependence on $\Xcal_0$ for notation simplicity. 

We first examine the effect of $\blambda$ selection on the quality of the mvBSC clustering. Table \ref{tb: compare to oracle} summarizes the mean and the standard deviation of the clustering accuracy and NMI score for the mvBSC clustering with $\blambda$ selected via mvBSC$_q$, mvBSC$\subSNR$ and the oracle method under the five network structures (M1)--(M5) with $K=25, \sigma_1=0.4, \sigma_2=0.6$. First, we note that both mvBSC$_q$ and mvBSC$\subSNR$ have comparable clustering performance to that of the mvBSC trained with oracle $\blambda$ across all settings. Although $\alpha_1 \ne \alpha_2$, selecting $\blambda$ based on the simple mvBSC$\subSNR$ appears to result in clustering with near identical performance as that of mvBSC$_q$. These results suggest that the proposed procedure for selecting $\blambda$ is indeed near optimal and the simple mvBSC$\subSNR$ works well when the views are reasonably similar. 
\begin{table}[htbp]
	\resizebox{0.9\textwidth}{!}{
		\begin{minipage}{\textwidth}	
			\begin{center}
				\begin{tabular}{c|c|c|c|c|c}
					\hline \hline
					\multirow{2}{*}{model} & \multirow{2}{*}{method} & \multicolumn{2}{c}{Accuracy} & \multicolumn{2}{c}{NMI score} \\ \cline{3-4} \cline{5-6} 
					& & mean & sd & mean & sd \\ \hline
					\multirow{3}{*}{M1} & oracle & 0.966 & 0.0169 & 0.989 & 0.0046 \\\cline{2-6}
					& mvBSC$_q$ & 0.954 & 0.0255 & 0.985 & 0.0065 \\ \cline{2-6}
					& mvBSC$\subSNR$ & 0.952 & 0.0255 & 0.984 & 0.0068 \\ \hline
					\multirow{3}{*}{M2} & oracle & 0.968 & 0.0145 & 0.988 & 0.0053 \\\cline{2-6}
					& mvBSC$_q$ & 0.945 & 0.0246 & 0.983 & 0.0061 \\ \cline{2-6}
					& mvBSC$\subSNR$ & 0.943 & 0.0272 & 0.983 & 0.0070 \\ \hline
					\multirow{3}{*}{M3} & oracle & 0.968 & 0.0121 & 0.989 & 0.0036 \\\cline{2-6}
					& mvBSC$_q$ & 0.947 & 0.0248 & 0.984 & 0.0063 \\ \cline{2-6}
					& mvBSC$\subSNR$ & 0.947 & 0.0257 & 0.984 & 0.0065 \\ \hline
					\multirow{3}{*}{M4} & oracle & 0.871 & 0.0272 &0.948 & 0.0112 \\\cline{2-6}
					& mvBSC$_q$ & 0.826 & 0.0356 & 0.936 & 0.0100 \\ \cline{2-6}
					& mvBSC$\subSNR$ & 0.822 & 0.0367 & 0.936 & 0.0101 \\ \hline
					\multirow{3}{*}{M5} & oracle & 0.734 & 0.0213 & 0.869 & 0.0156 \\\cline{2-6}
					& mvBSC$_q$ & 0.680 & 0.0344 & 0.857 & 0.0173 \\ \cline{2-6}
					& mvBSC$\subSNR$ & 0.671 & 0.0355 & 0.856 & 0.0182 \\ \hline			
				\end{tabular}
			\end{center}
			\caption{The average clustering accuracy and NMI of mvBSC procedures under five different models with a medium noise level, $K=25$ with optimal $\blambda$ selected based on (\ref{eq: best.q}) (mvBSC$_q$), the the empirical version of (\ref{eq: snr}) (mvBSC$\subSNR$) and the oracle obtained by minimizing the empirical $|\Mcal_{\blambda^*}|$.}
			\label{tb: compare to oracle}
	\end{minipage}}
\end{table}

\begin{figure}[htbp]
	\begin{center}
		\subfloat[small noise]{\includegraphics[width=5cm,height=7cm]{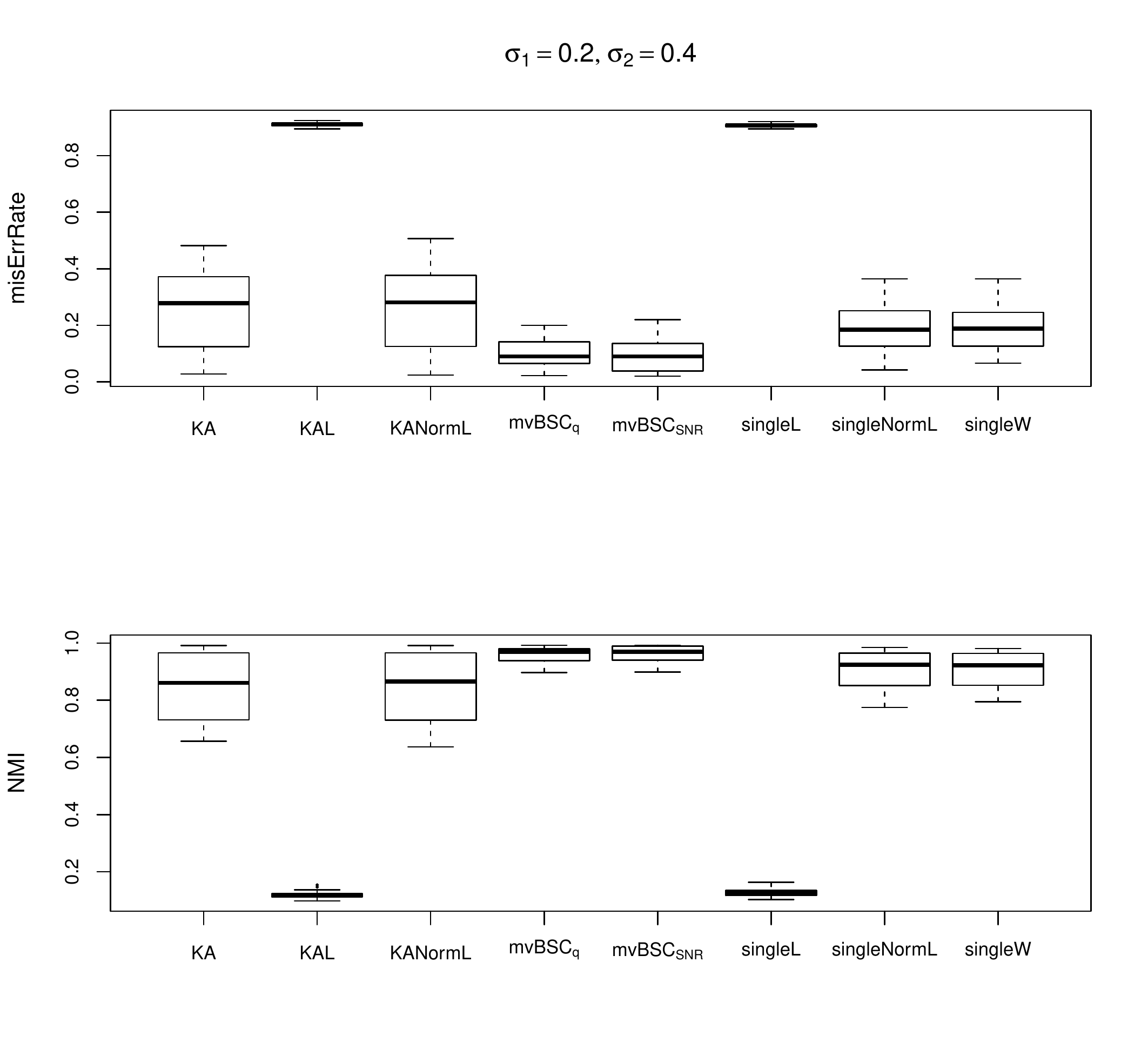}} \quad
		\subfloat[medium  noise]{\includegraphics[width=5cm,height=7cm]{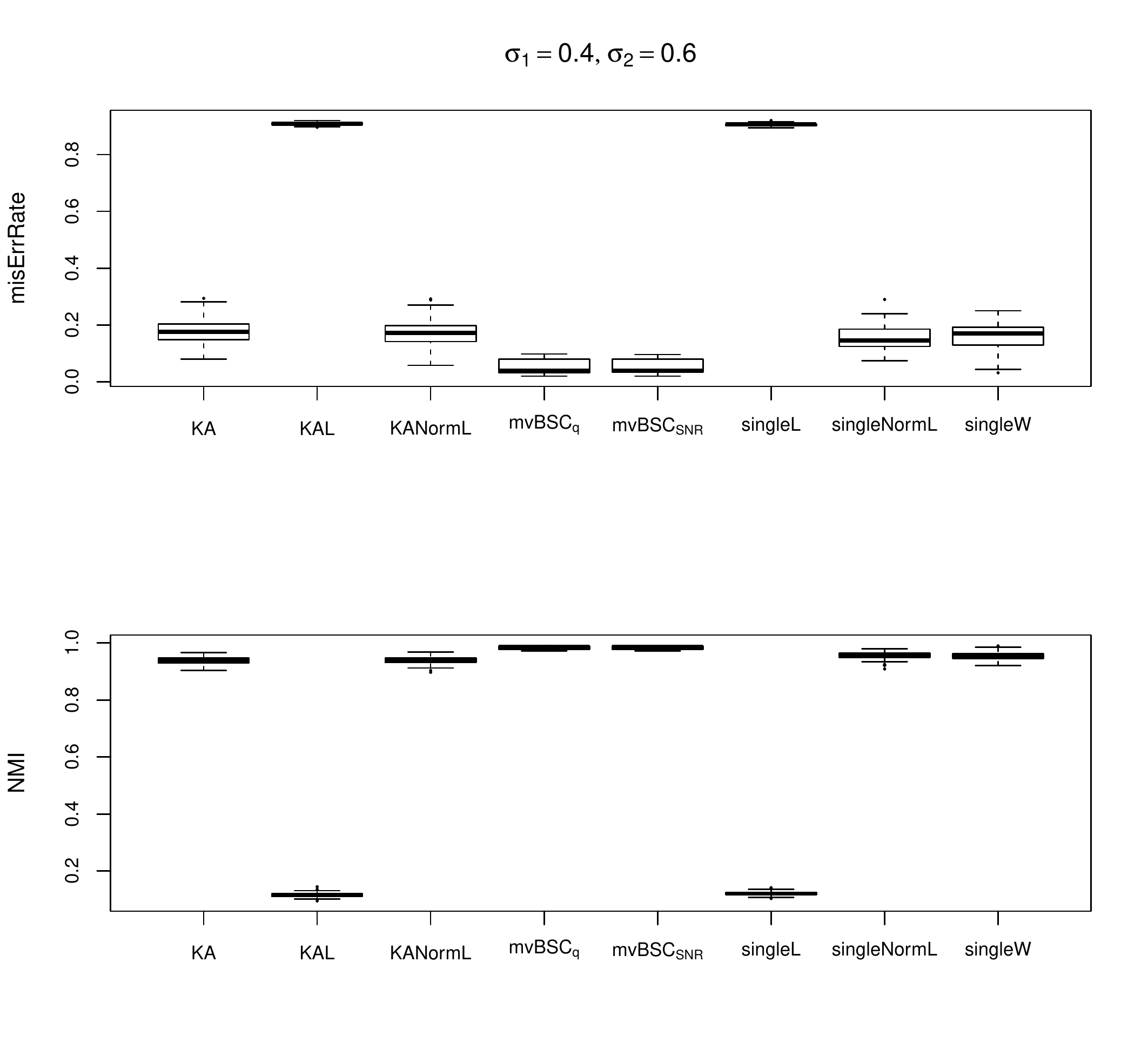}} \quad
		\subfloat[large noise]{\includegraphics[width=5cm,height=7cm]{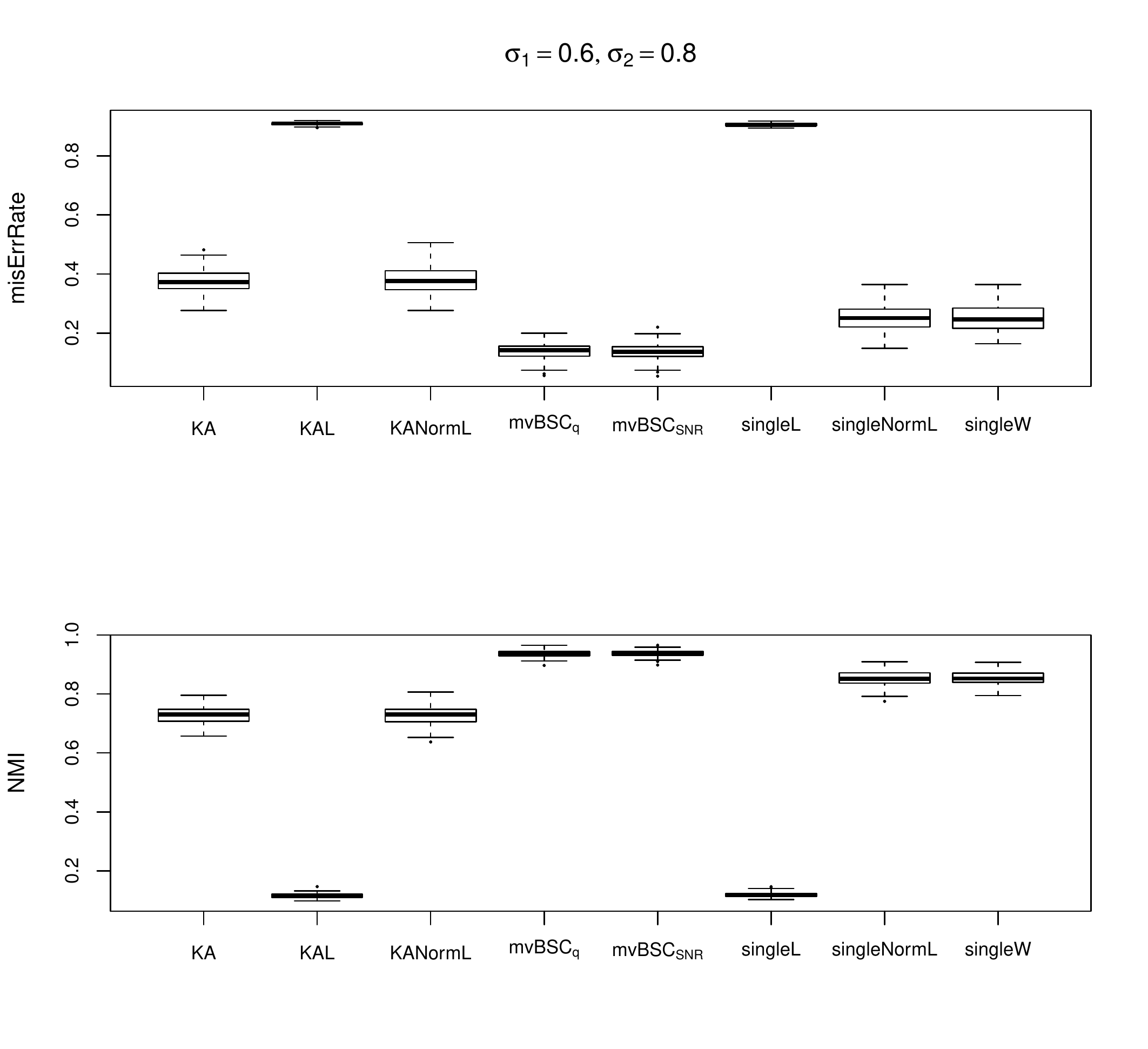}}
		\caption{Boxplots of clustering accuracy and NMI of different clustering methods for (M3) with $K = 25$ and three different noise levels: (i) mvBSC$_q$: mvBSC using (\ref{eq: best.q}), (ii) mvBSC$\subSNR$: mvBSC using (\ref{eq: snr}), (iii) clustering using the kernel addition matrix $\Wbb^+$ (KA); (iv) clustering with the Laplacian of the $\Wbb^+$ (KAL); (v) clustering using the normalized Laplacian of $\Wbb^+$ (normKAL); (vi) $\Wbb^1$ (singleW), (vii)  clustering with Laplacian of a single $\Wbb$ (singleL); and (viii) clustering with the normalized Laplacian of $\Wbb^1$.}
		\label{fig: compare to baseline three noise levels in M3}
	\end{center}
\end{figure}

\begin{figure}[htbp]
	\begin{center}
		\subfloat[low $K/n$]{\includegraphics[width=5cm,height=7cm]{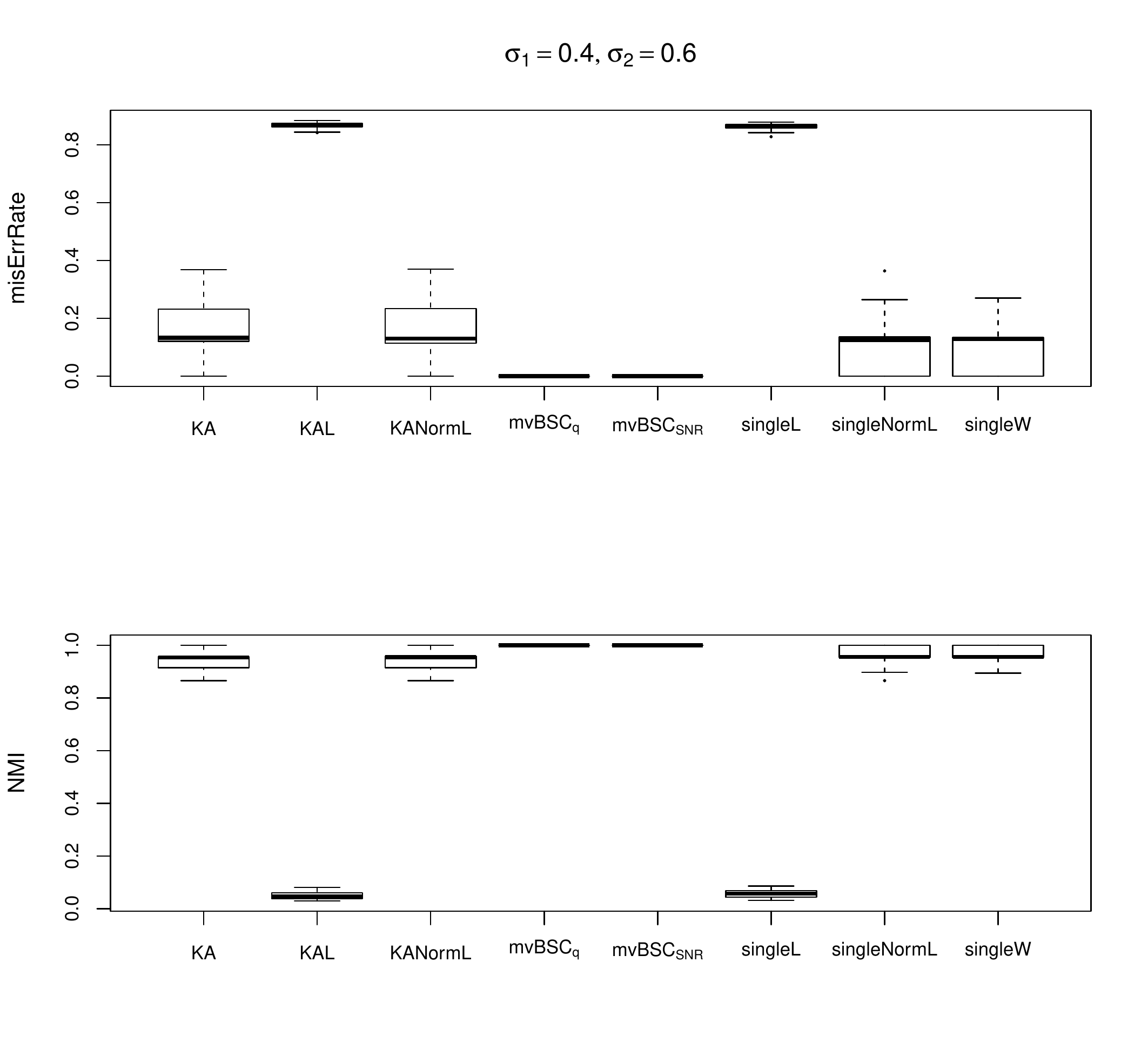}} \quad
		\subfloat[medium  $K/n$]{\includegraphics[width=5cm,height=7cm]{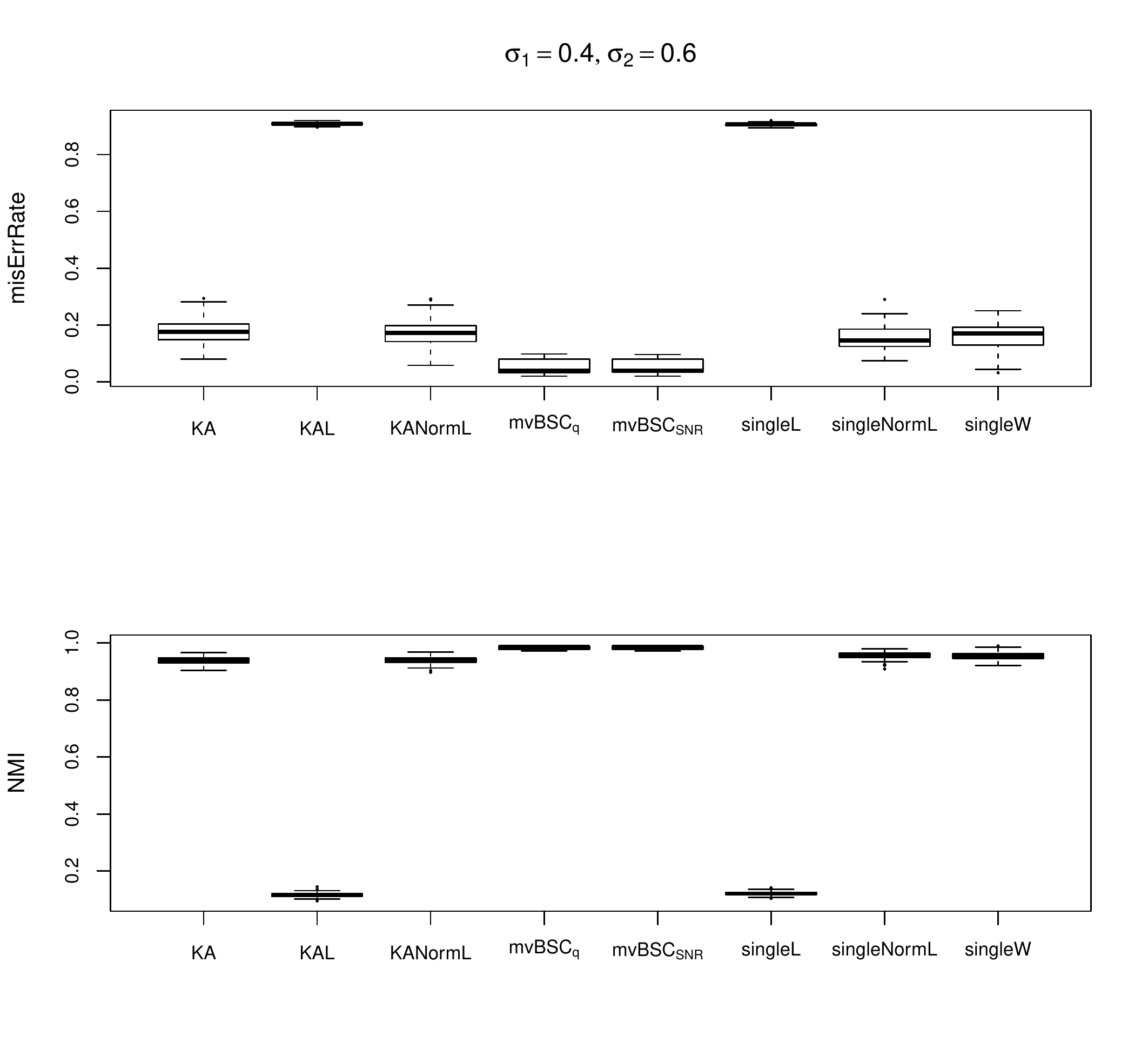}} \quad
		\subfloat[large $K/n$]{\includegraphics[width=5cm,height=7cm]{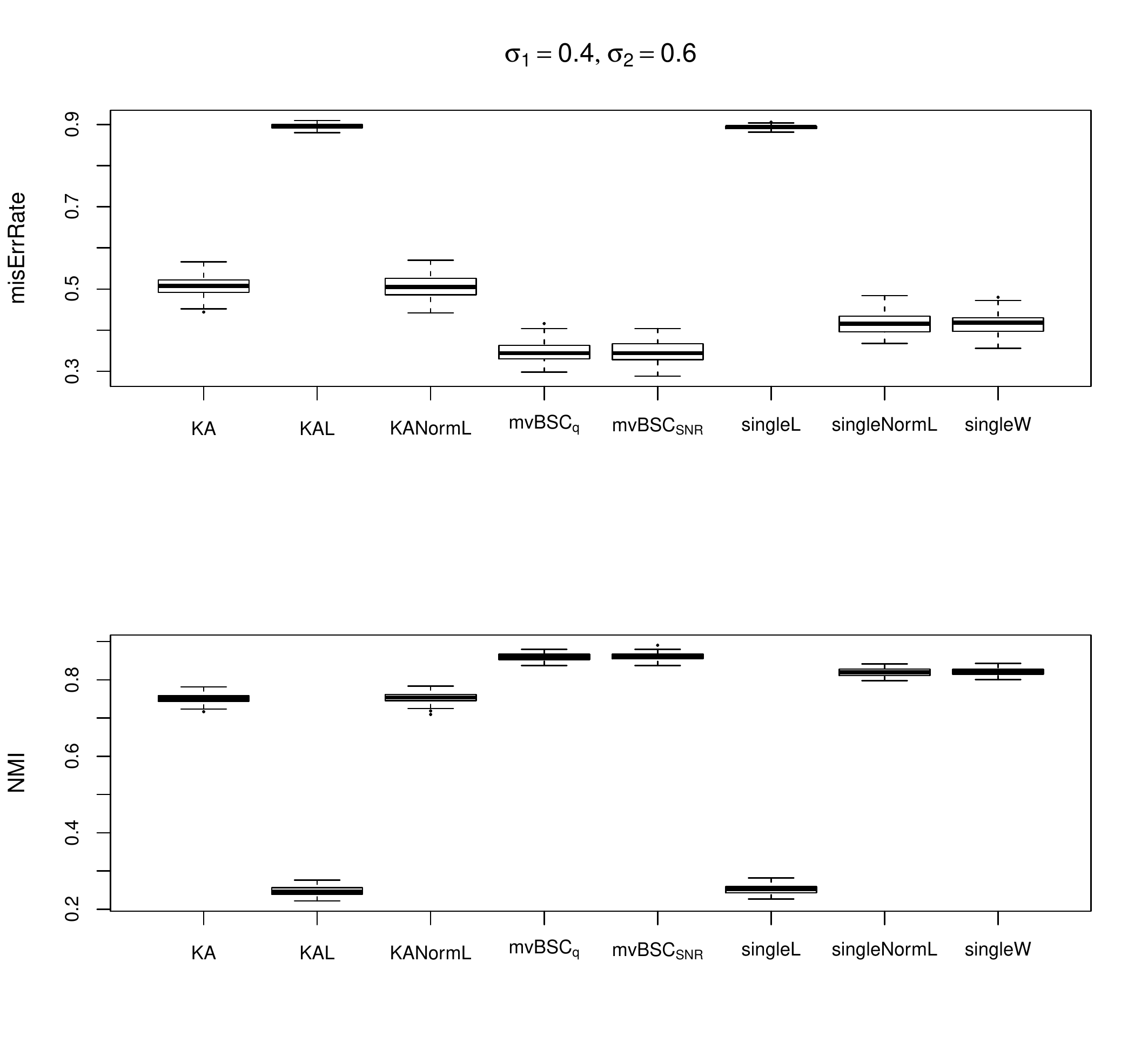}}
		\caption{Boxplots of mis-clustered error rate and NMI score  over multiple methods in comparison by three different $K/n$ network structures. mvBSC$_q$: mvBSC using (\ref{eq: best.q}), mvBSC$\subSNR$: mvBSC using (\ref{eq: snr}), KA: the kernel addition matrix, KAL: Laplacian of the kernel addition matrix, normKAL: the normalized Laplacian of the kernel addition matrix, singleW: a single $\Wbb$, singleL: Laplacian of a single $\Wbb$, singleNormL: the normalized Laplacian of a single $\Wbb$.}
		\label{fig: compare to baseline three K/n levels in M3}
	\end{center}
\end{figure}

We next compare the performance of mvBSC$_q$ and mvBSC$\subSNR$ to the aforementioned alternative spectral clustering procedures. In Figure \ref{fig: compare to baseline three noise levels in M3}, we show the clustering accuracy  and NMI for different clustering methods under (M3) with $K = 25$ and different noise levels for $\Wbb^s$. For conciseness of the presentation, for the methods based on a single view, we only report the maximum accuracy and NMI of the two views.  It is easy to see the normalized Laplacian always performs better than its unnormalized counterpart and in fact the unnormalized version fails in all scenarios. The KA clustering with $\Wbb^+$ performs even worse than the clustering with the single view $\Wbb^s$, suggesting that a naive aggregation of multiple sources of information could have detrimental effect on the clustering due to the heterogeneity in the underlying $\Omega^s$. Our mvBSC method is consistently better than all competing methods in terms of both average and spread across all noise levels and the advantage is even more apparent as noise level increases. Figure \ref{fig: compare to baseline three K/n levels in M3} shows how the performances of different methods change over different level of $K$ under the medium noise level setting. As $K$ increases, the clustering becomes more challenging. As a result, the clustering accuracy and NMI decrease substantially for most competing methods but only slightly for the mvBSC method. Thus, the larger noise level and $K$, the more advantage the mvBSC approach showcases over other methods. 

\section{Grouping ICD9 Codes with mvBSC}
\label{sec: real data}

\subsection{Motivation}
The International Classification of Disease, 9th edition (ICD9) coding system, containing over 14,000 codes, is a widely adopted mechanism for billing. Recording a full spectrum of diagnoses and procedure information in the electronic health records (EHR), the ICD9 coding system is a valuable resource for various types of biomedical research. However, designed for billing and administrative functions, individual ICD9 codes tend to be too specific to be directly used as disease phenotypes. Many codes indeed describe the same disease and only differ in details such as the affected anatomical areas. For clinical and genetic studies, it is thus often desirable to collapse detailed codes into clinically relevant groups. To address such a need, \cite{denny2010phewas,denny2013systematic} manually curated grouping information to allow for more efficient representation of disease phenotypes recorded in the EHR. The grouping has been successfully used to perform phenome-wide association studies (PheWAS). Despite a valuable asset, this manual curation approach has major limitations including lack of scalability, portability and susceptible to subjective bias. With the adoption of ICD10 codes in recent years, a substantial human effort will be required to manually update the grouping to include both ICD9 and ICD10 codes, signifying the need of a data-driven approach. 

\subsection{Data Sources and Model Set-up}
\begin{figure}[htbp]
	\begin{center}
		\subfloat[claim]{\includegraphics[width=5cm,height=5cm]{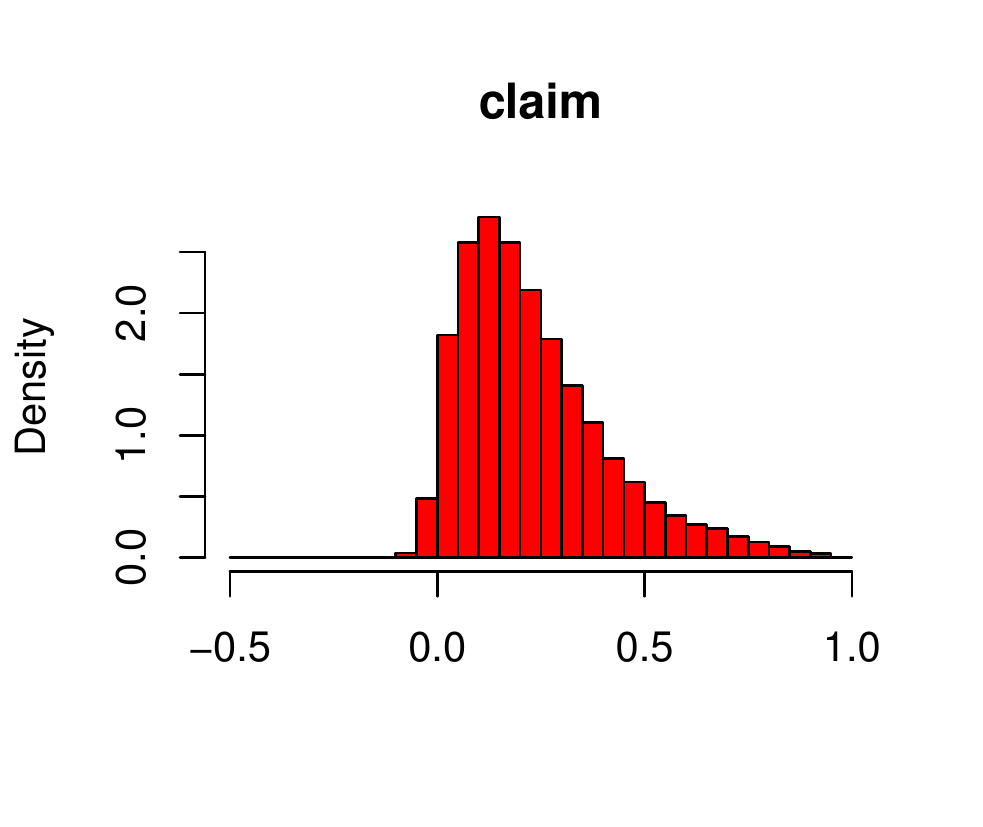}} \quad
		\subfloat[va]{\includegraphics[width=5cm,height=5cm]{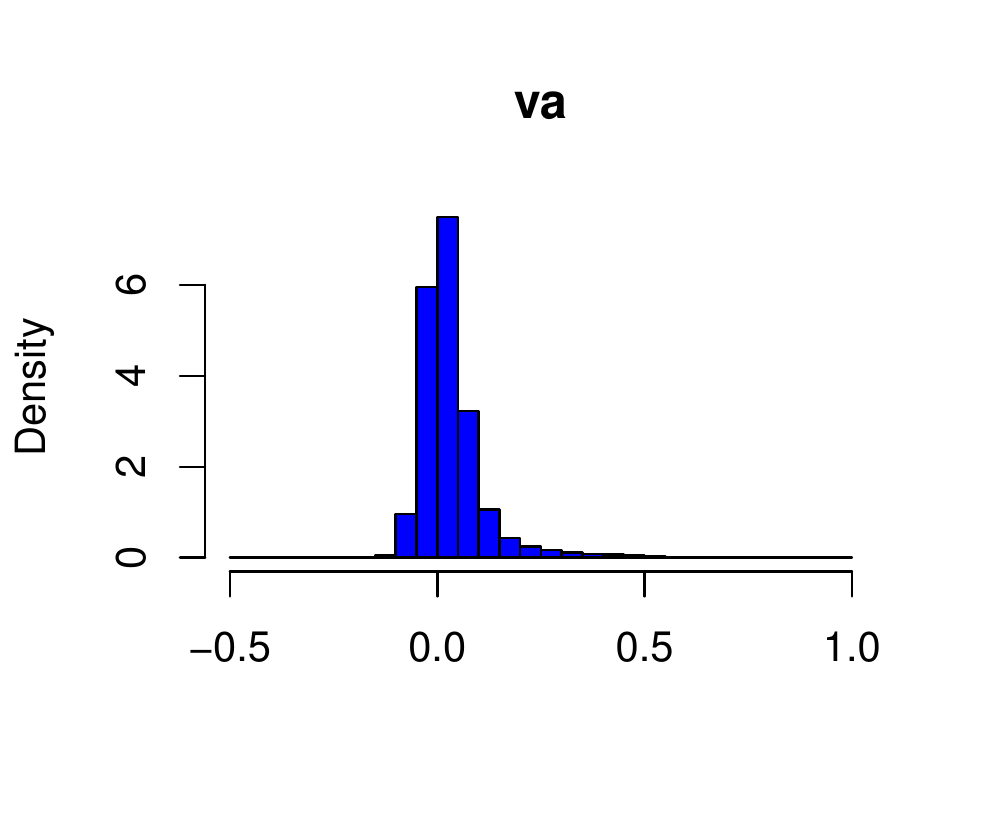}} \quad
		\subfloat[biobank]{\includegraphics[width=5cm,height=5cm]{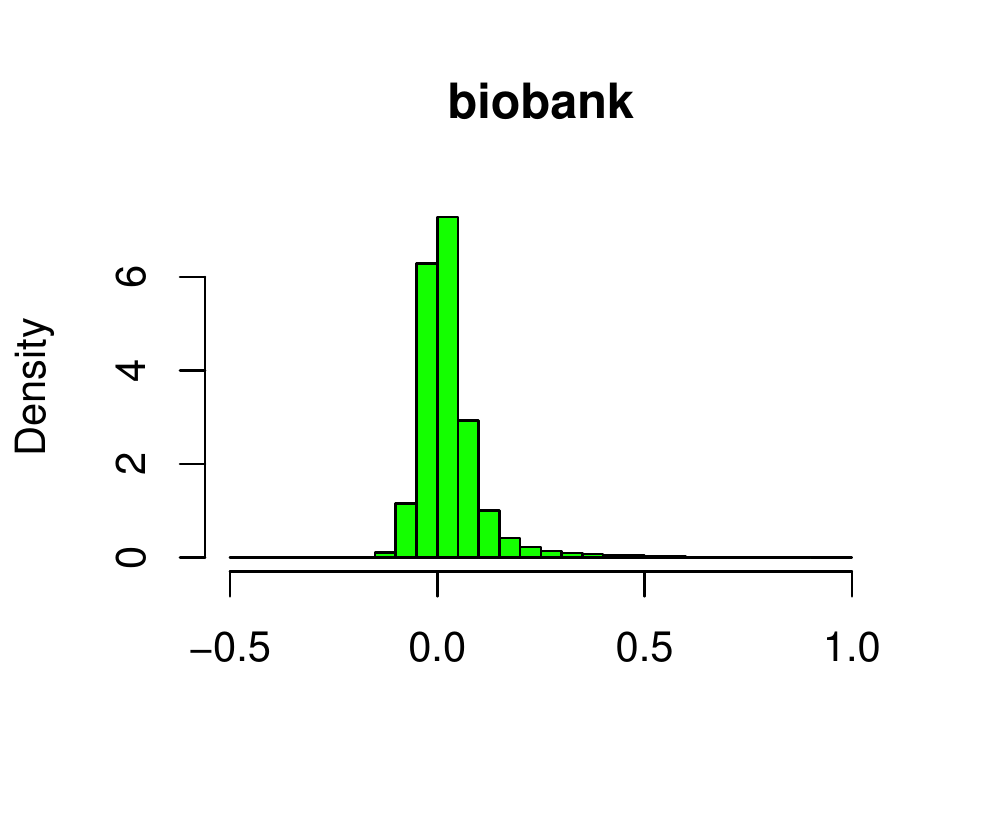}} \\
		\subfloat[claim]{\includegraphics[width=5cm,height=5cm]{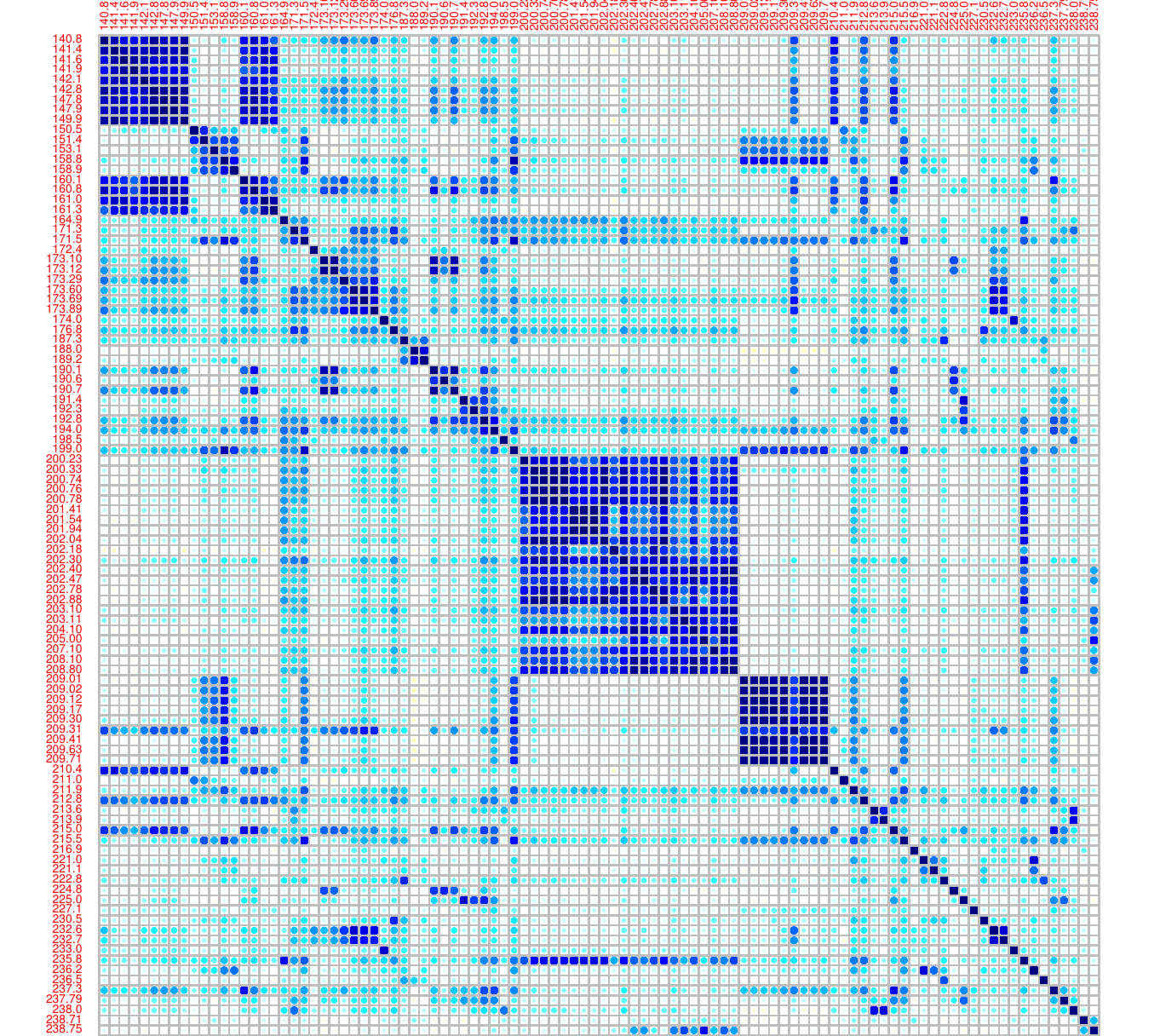}} \quad
		\subfloat[va]{\includegraphics[width=5cm,height=5cm]{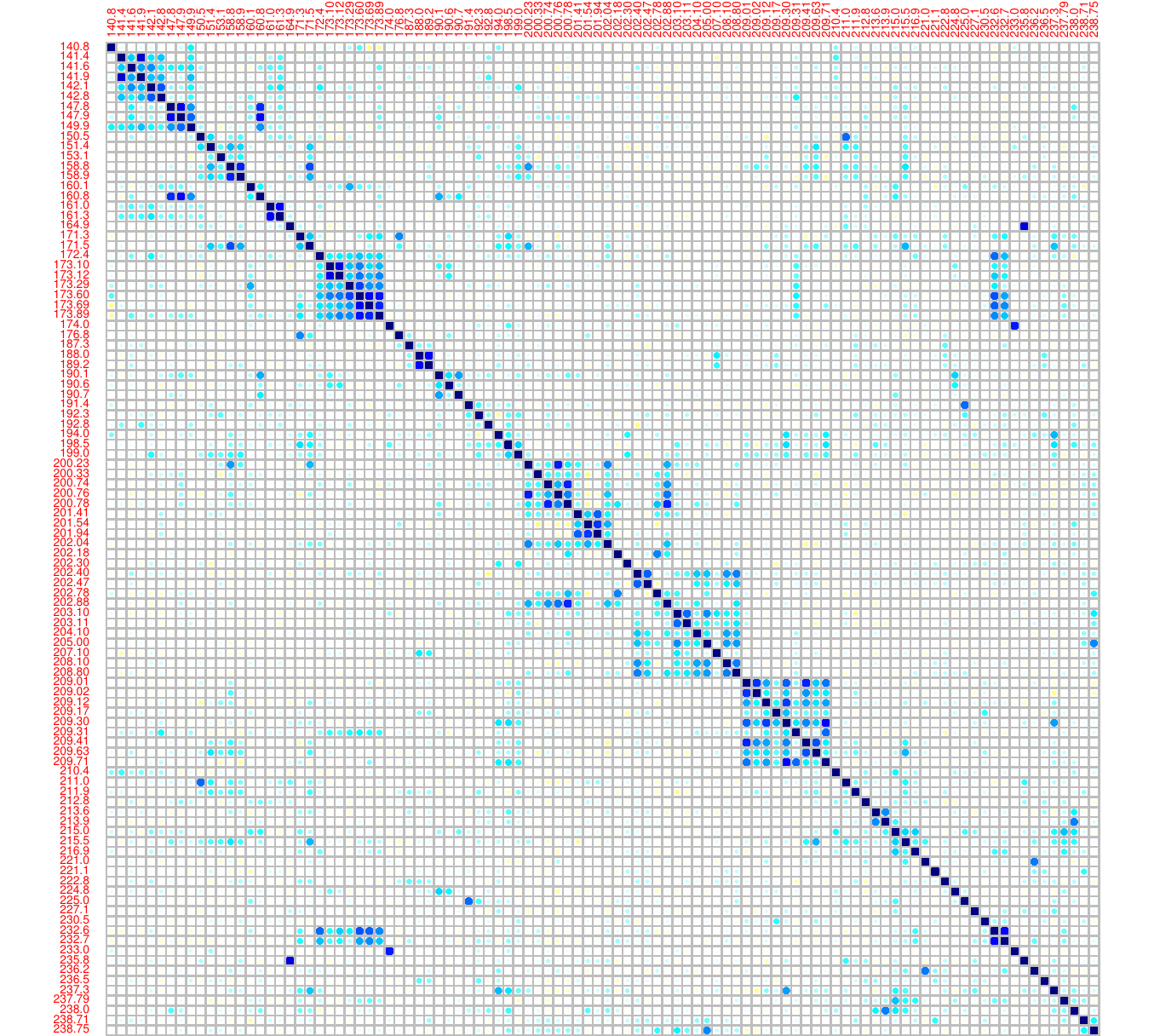}} \quad
		\subfloat[biobank]{\includegraphics[width=5cm,height=5cm]{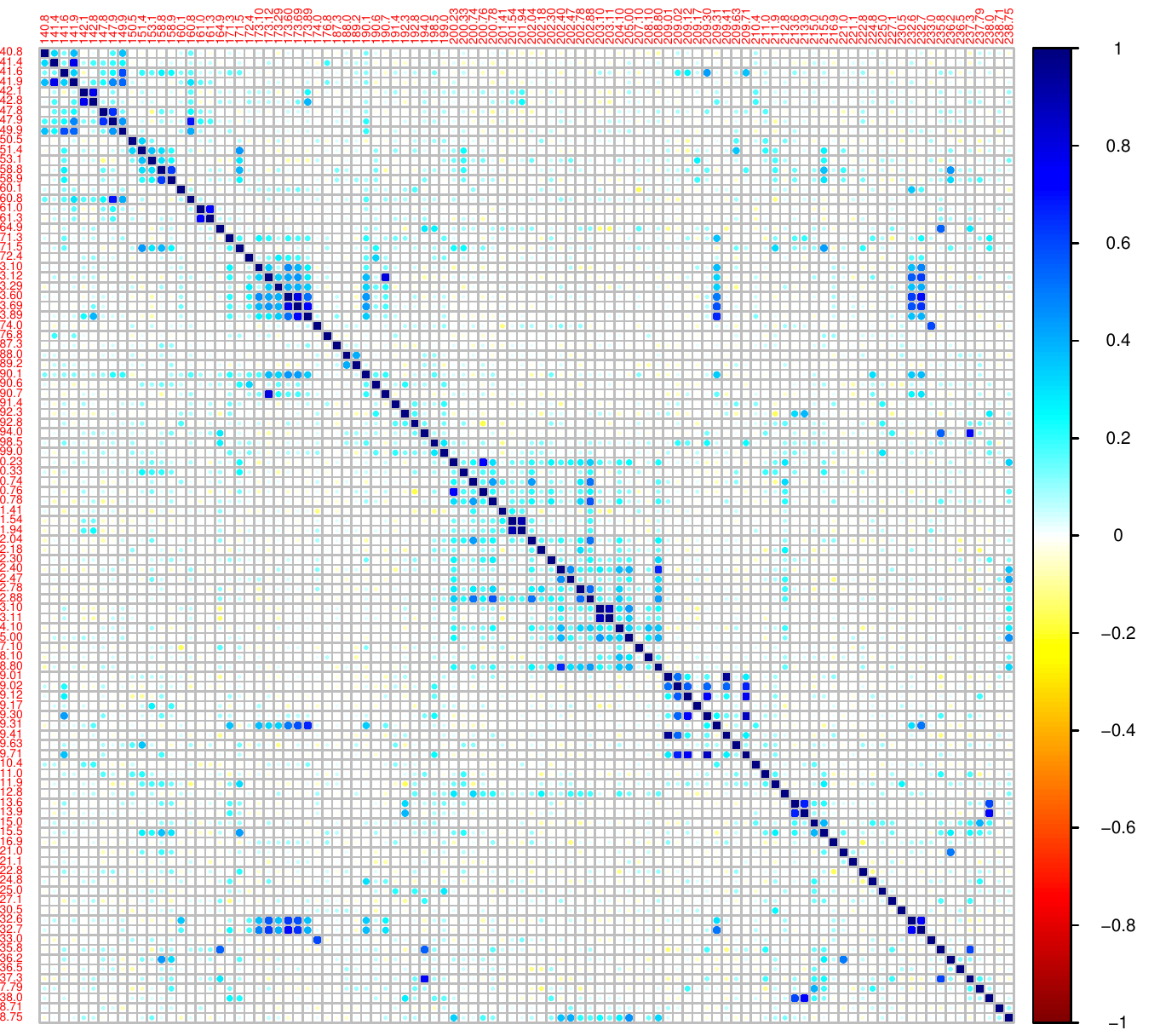}} 
	\end{center}
	\caption{Raw data summary: the top panel displays the density histogram of each cosine similarity matrix. 
	The bottom panel corresponds to each cosine similarity matrix restricted on a common set of codes.}
	\label{fig: summarize raw data}
\end{figure}

To employ the proposed mvBSC algorithm, three similarity matrices $\{\Wbb^s, s = 1, 2, 3\}$ were obtained for all ICD9 codes from three different healthcare systems
including a large insurance claim database (Claim), the Veteran Health Administration (VHA) and Partner's Healthcare systems (PHS). 
Here, $W_{ij}^s$ represents the cosine similarity score of the semantic vectors for ICD9 codes $v_i$ and $v_j$ from the $s$th data source, 
Within each healthcare system, the semantic vectors were obtained by fitting a {\em word2vec} algorithm  \citep{mikolov2013efficient} to a co-occurrence table that records the frequency of a code pair co-occuring within a 30-day time window. Two main factors contribute to the heterogeneity across the three data sources. First, the sample sizes are significantly different stretching from $\sim 45$ million for Claim down to 1 million for VHA and further reducing to $\sim 60,000$ for Partner's Biobank. Second, the underlying patient populations vary substantially. Specifically, Claim covers a full nationwide spectrum of subjects, whereas VHA solely targets the veteran population and PHS primarily consists of tertiary hospitals enriched for patients with more complex and severe diseases. Such heterogeneities signify the need for an unbiased approach to optimally combine information from these sources, which can be also easily checked in Figure \ref{fig: summarize raw data} that gives a summary of the raw data on the cosine similarity matrices. The top panel displays the density histogram of each cosine similarity matrix, which supports the sparseness in each cosine similarity matrix, even though the sparseness pattern in claim is not as apparent as the other two.  The bottom panel is a snapshot of each cosine similarity matrix restricted on a common set of codes. The darkness of dots indicates the magnitude of the corresponding cosine similarity. Most large entries locate near the diagonal, suggesting the appropriateness of banding the three similarity matrices.
\def\Ncal{\mathcal{N}}
\def\Nscr{\mathscr{N}}

\begin{figure}[htbp]
	\begin{center}
		\includegraphics[width=10cm,height=1.5cm]{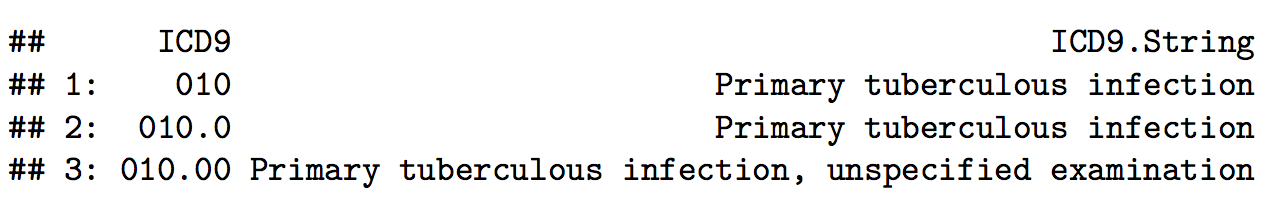}
		\caption{Ambiguity example in ICD9 coding system}
		\label{fig: ambiguity}
	\end{center}
\end{figure}

For an ICD9 code $v_i$, we let $d_{ij} = |\Ncal(v_i) - \Ncal(v_j)| + \eta I\{v_i \ne v_j, \Ncal(v_i) = \Ncal(v_j)\}$, where $\Ncal(\cdot)$ maps a character string to its numeric form and $\eta$ is a small constant chosen such as $0.005$. For example, $\Ncal(``001.1") = 1.1$. As a consequence, the vertex set $\Vcal$ can be ordered in the sense that $\Ncal(v_i) < \Ncal(v_j)$ if and only if $i<j$. The additional term involving a small constant $\eta$ is included to distinguish the case of $v_i = v_j$ versus the rare cases when $\Ncal(v_i) = \Ncal(v_j)$ but $v_i\ne v_j$ (see for example Figure \ref{fig: ambiguity}). 

 In this application, the existing manually curated PheWAS groups can also serve as silver-standard labels to provide a guidance in search for appropriate choices of the banding parameter $h_s$ and the weight parameter $\lambda_s$. All parameters were tuned in a grid search manner with the best corresponding to the highest NMI score. To choose a proper $K$, considering the number of PheWAS groups($K_{phewas}$) is already a good estimate, we scanned through its neighborhood ($0.8K_{phewas} \sim 1.2K_{phewas}$) and picked the one with the highest NMI score for subsequent analysis($K_{use}$). With a given $K$, we performed clustering using the proposed mvBSC$\subSNR$ procedure along with the mvBSC procedure with $\blambda$ selected to empirically maximize the NMI (mvBSC$_{\scriptscriptstyle \sf maxNMI}$). Results based on mvBSC$_q$ are omitted here since three banding parameters are very close to each other which yield very similar results to mvBSC$\subSNR$.

\subsection{Results}
\begin{figure}[htbp]
	\begin{center}
		\subfloat[neurological, mvBSC$\subSNR$]{\includegraphics[scale=0.3]{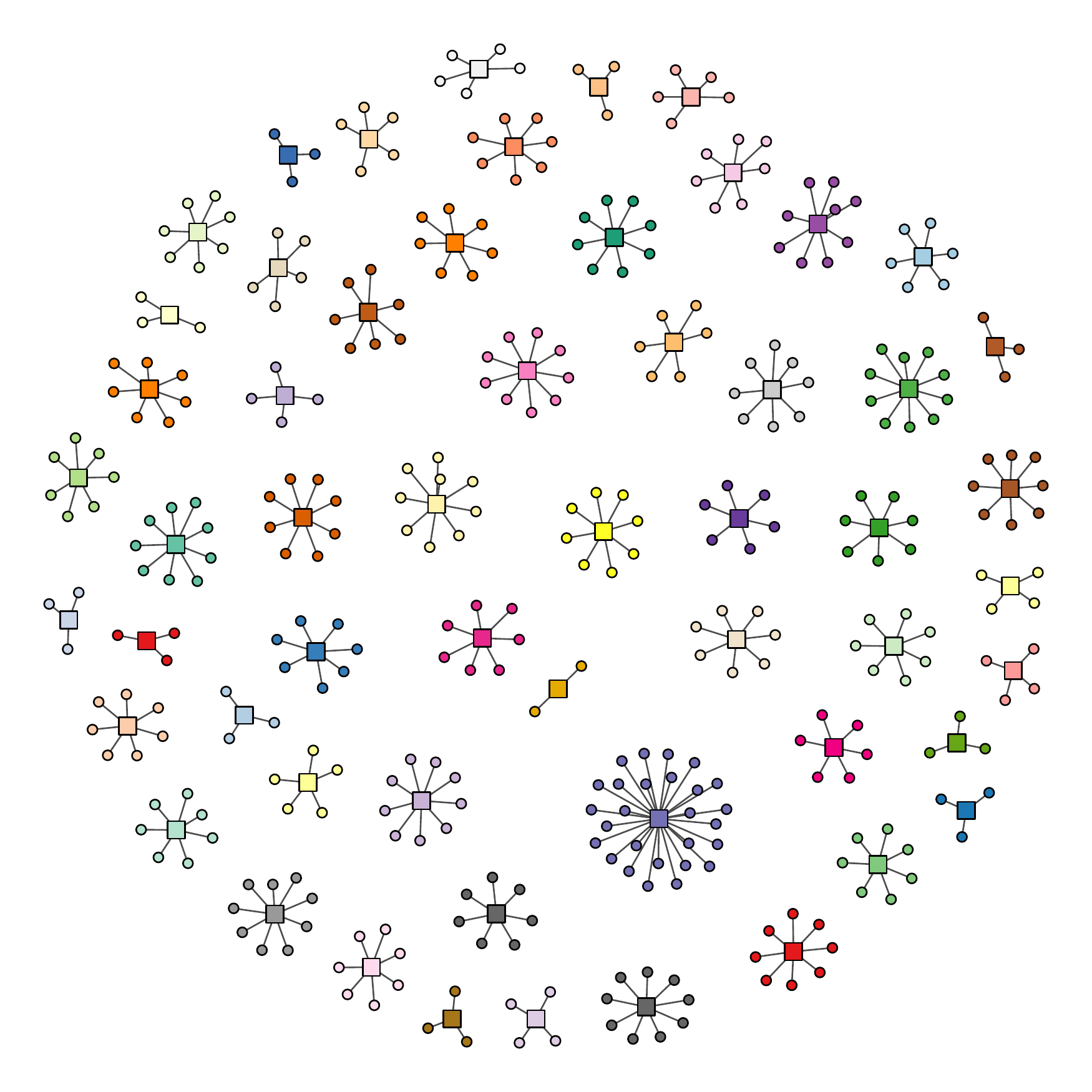}} \quad
		\subfloat[neurological, PheWAS ]{\includegraphics[scale=0.3]{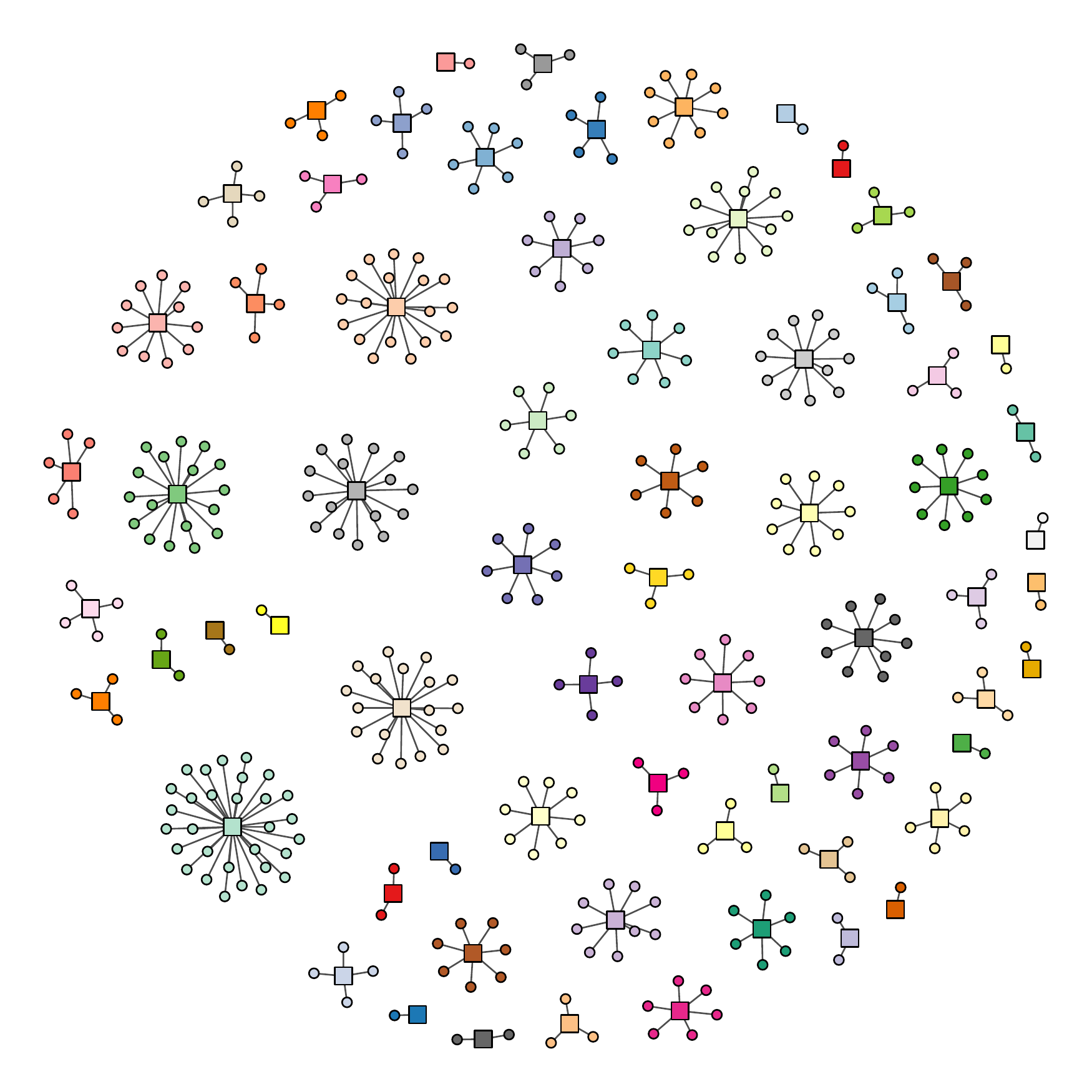}} \\
		\subfloat[musculoskeletal, mvBSC$\subSNR$]{\includegraphics[scale=0.3]{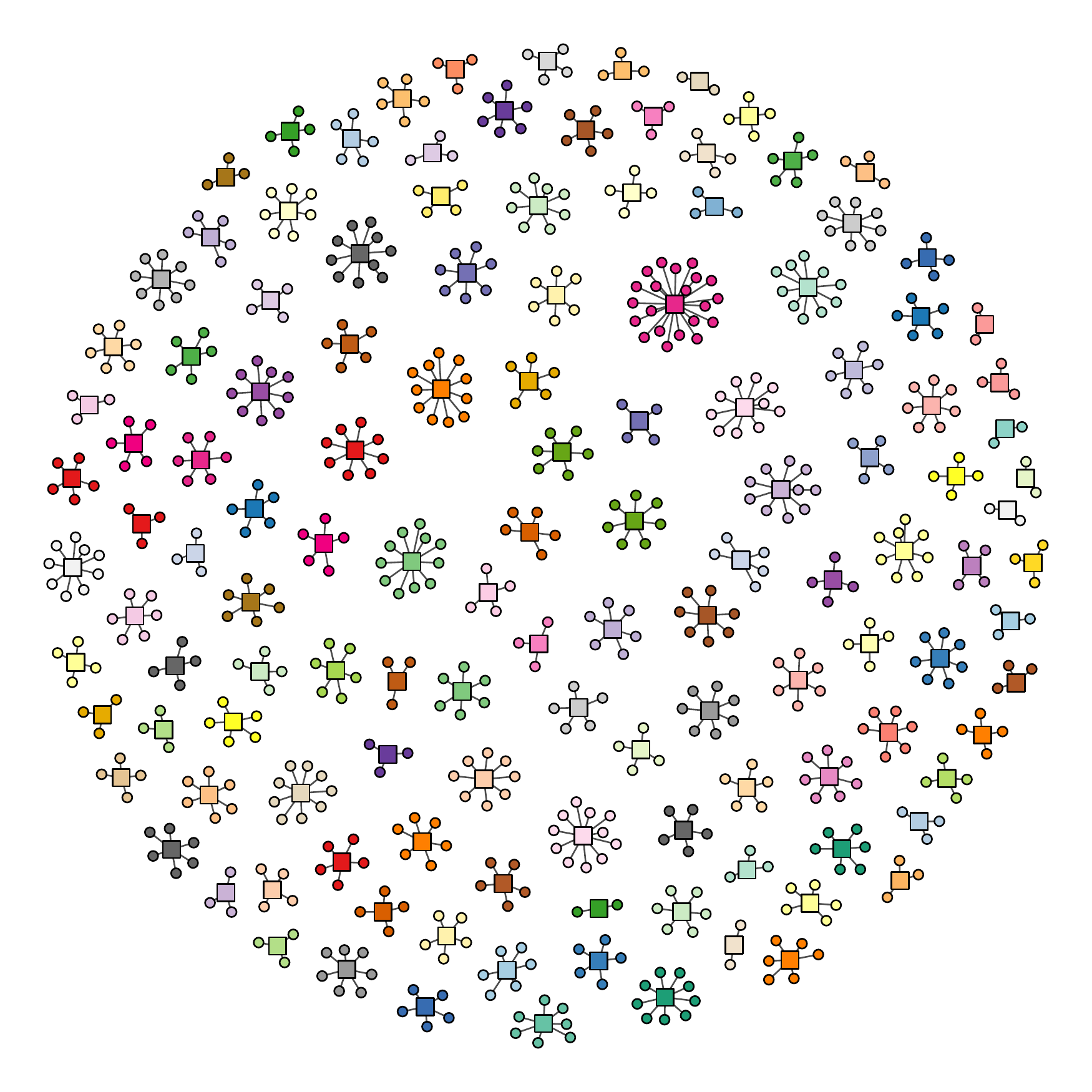}} \quad
		\subfloat[musculoskeletal, PheWAS ]{\includegraphics[scale=0.3]{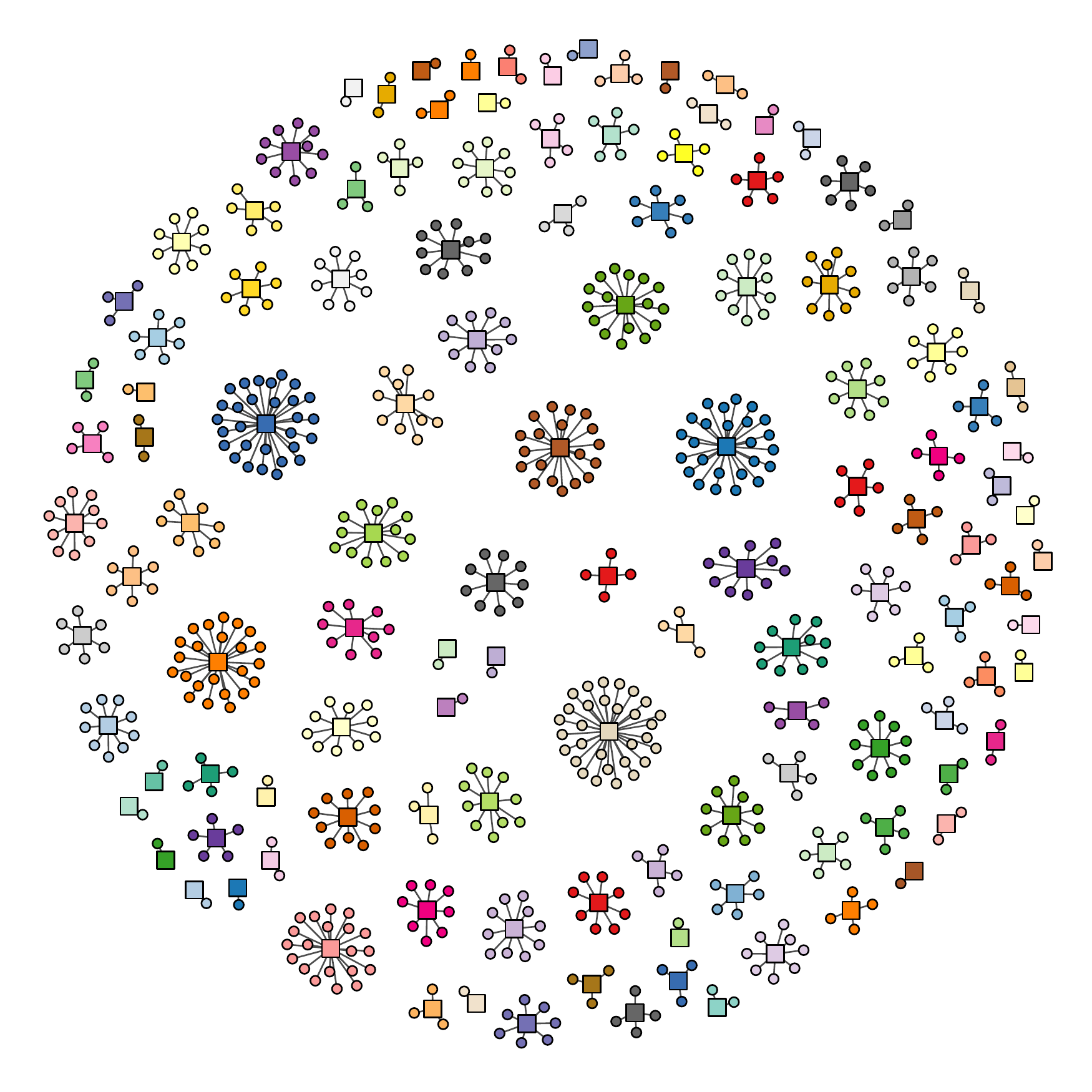}}         
		\caption{Clustering structure comparison. Squares represent cluster nodes and circles represent ICD9 codes.}		
		\label{fig: full clustering comparison}
	\end{center}
\end{figure}

For illustration purposes, in this paper we only present results focusing on the following four categories--neoplasms, neurological, musculoskeletal and sense organs -- whose grouping results draw particularly great interest in the current clinical studies. Table \ref{tb: NMI score} clearly shows that the proposed mvBSC algorithm performs well across four categories with high agreement with the existing PheWAS grouping. Beyond that, our proposed mvBSC algorithm has the advantage of being efficient, scalable, and adaptive to the evolving human knowledge as reflected in the observed data. The clustering with $\blambda$ selected via mvBSC$\subSNR$  also has similar performance as the optimal $\lambda$ selected to maximize the NMI. Figure \ref{fig: full clustering comparison} visually compares the the global clustering structure given by mvBSC$\subSNR$ and PheWAS on neurological and musculoskeletal category respectively, showing the power of mvBSC$\subSNR$ to mimic the global network structure. 
\begin{table}[htbp]
		\resizebox{0.9\textwidth}{!}{
		\begin{minipage}{\textwidth}
	\begin{center}
		\begin{tabular}{c|c|c|c|c|c}
			\hline \hline
			\multirow{2}{*}{Category} & \multirow{2}{*}{$n$} & \multirow{2}{*}{$K_{phewas}$} & \multirow{2}{*}{$K_{use}$} & \multicolumn{2}{c}{NMI score}\\	\cline{5-6} 
			& & & & mvBSC$_{\scriptscriptstyle \sf maxNMI}$  & mvBSC$\subSNR$\\ \hline
			neoplasms & 799 & 122 & 138 & 0.856 & 0.843\\ \hline
			neurological & 364 & 68 & 56 & 0.852 & 0.839 \\ \hline
			musculoskeletal & 675 & 124 & 128 & 0.834 & 0.790 \\ \hline
			sense organs & 639 & 119 & 132 & 0.862 & 0.859 \\ \hline
		\end{tabular}
		\caption{Table of NMI scores of the mvBSC procedure compared to PheWAS labels using two different choices of $\blambda$ (maximizing empirical NMI and SNR) across four different ICD categories, where $n$ is the total number of ICD9 codes within each category, $K_{phewas}$ is the total number of PheWAS groups and $K_{use}$ is the total number of groups used for final clustering by maximizing the NMI score.}
		\label{tb: NMI score}
	\end{center}
\end{minipage}}
\end{table} 

\begin{figure}[htbp]
	\begin{center}
		\includegraphics[width=7cm,height=4cm]{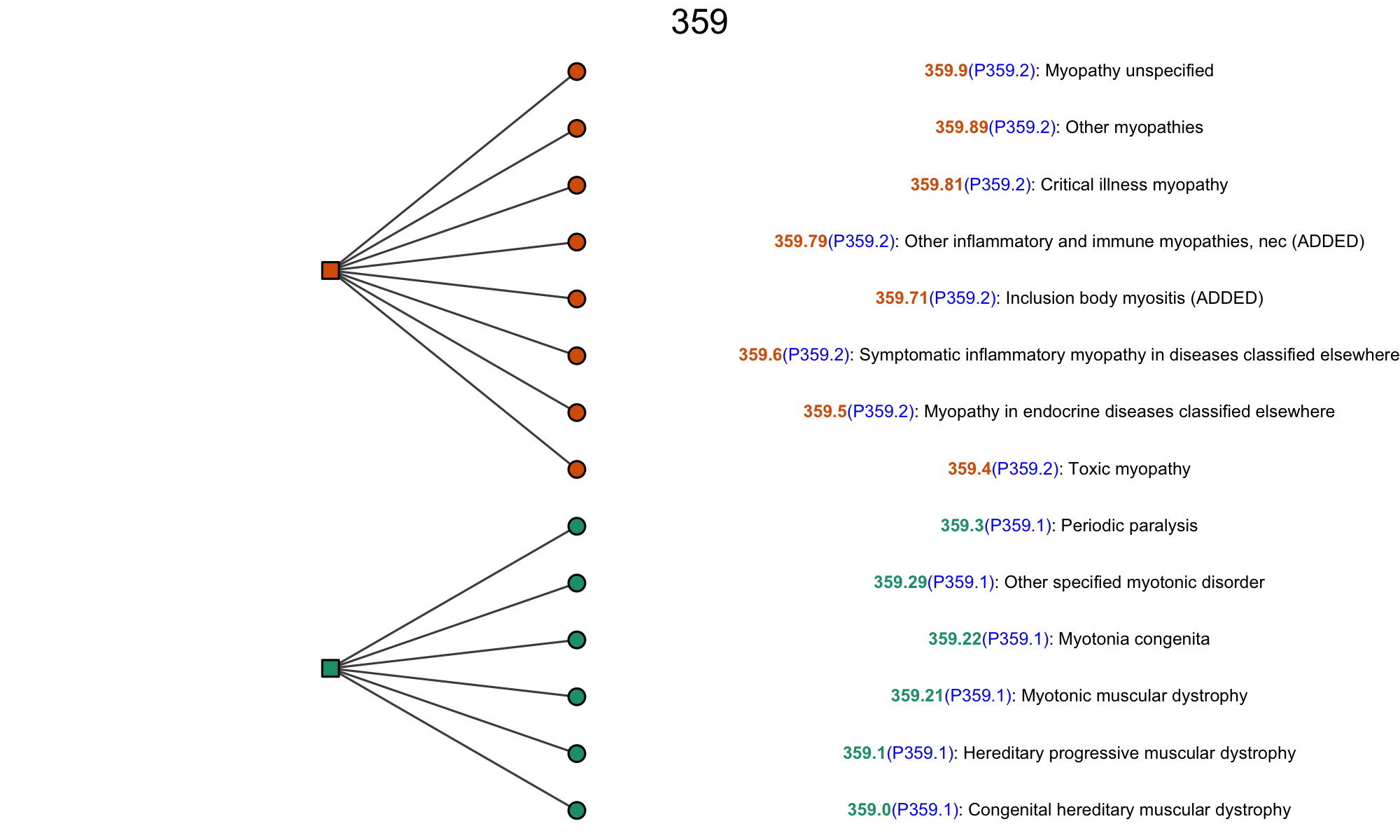}
		\caption{Grouping result for myopathy. Codes are colored by their membership. Codes in parenthesis colored in blue represent corresponding PheWAS codes.}
		\label{fig: 359}
	\end{center}
\end{figure}

\begin{figure}[htbp]
	\begin{center}
		\subfloat[mvBSC$\subSNR$]{\includegraphics[width=8cm,height=5cm]{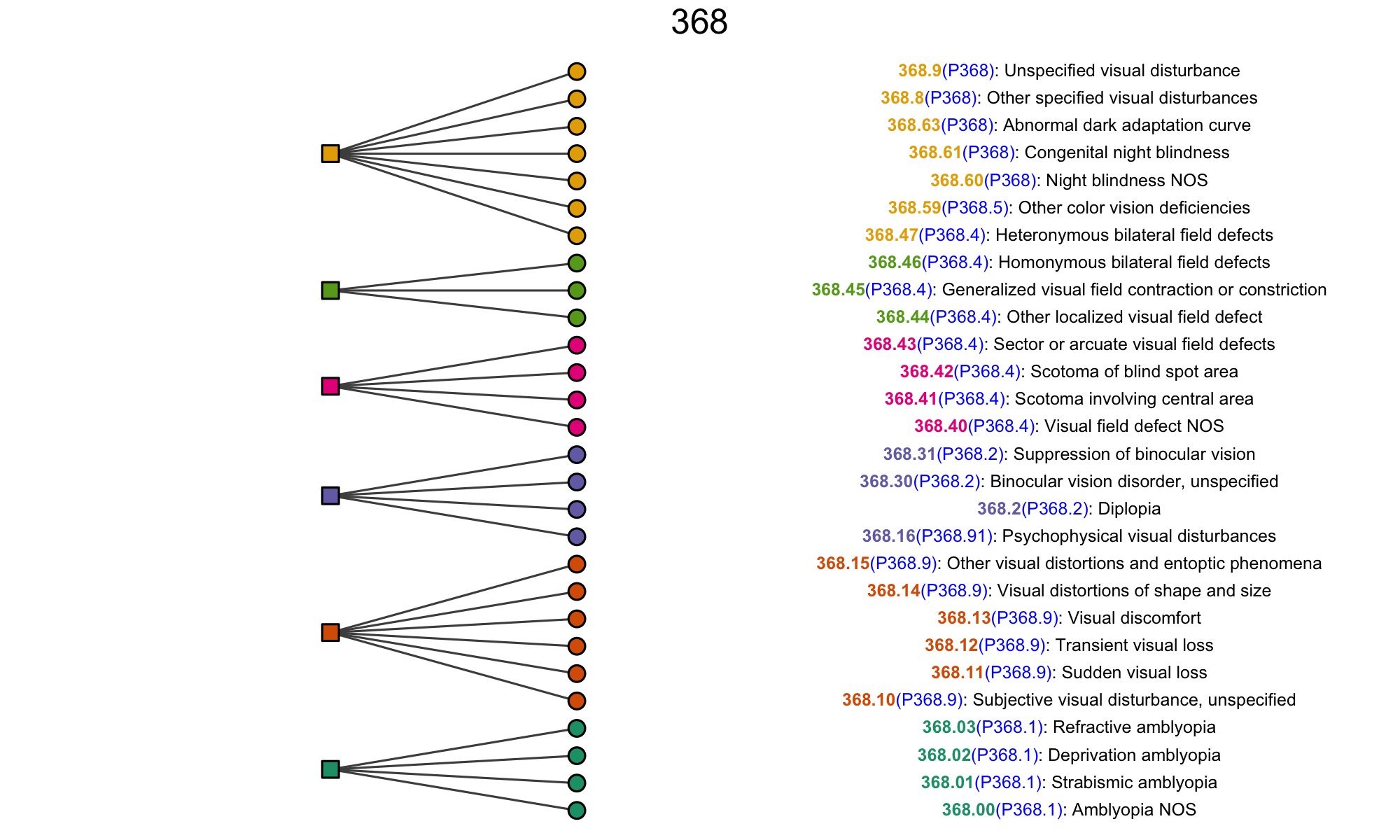}} \quad
		\subfloat[best grid]{\includegraphics[width=8cm,height=5cm]{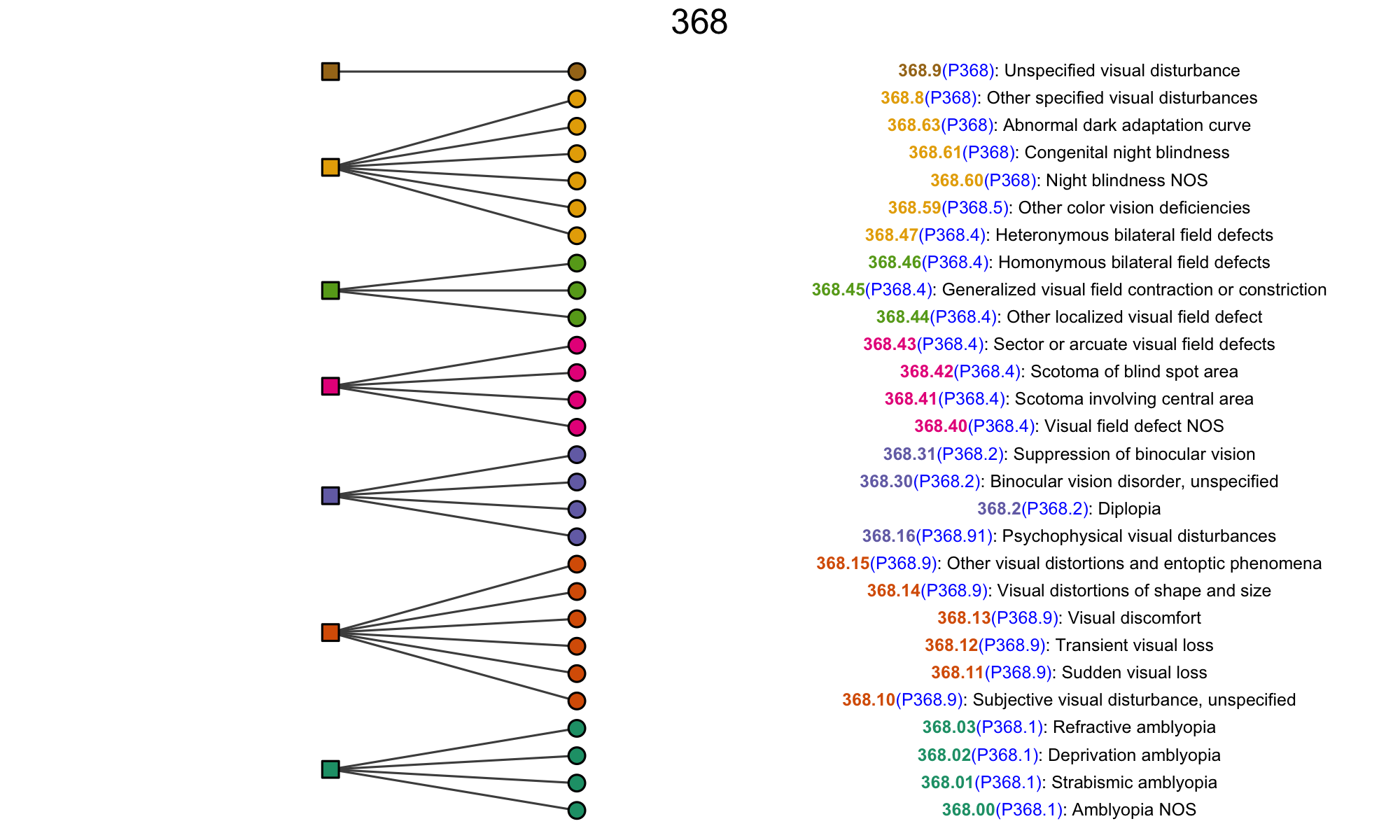}} \quad
		\caption{Grouping result for ICD9 . Codes are colored by their membership. Codes in parenthesis  colored in blue represent corresponding PheWAS codes.}
		\label{fig: 368}
	\end{center}
\end{figure}

\begin{figure}[htbp]
	\begin{center}
		\subfloat[711]{\includegraphics[width=8cm,height=7cm]{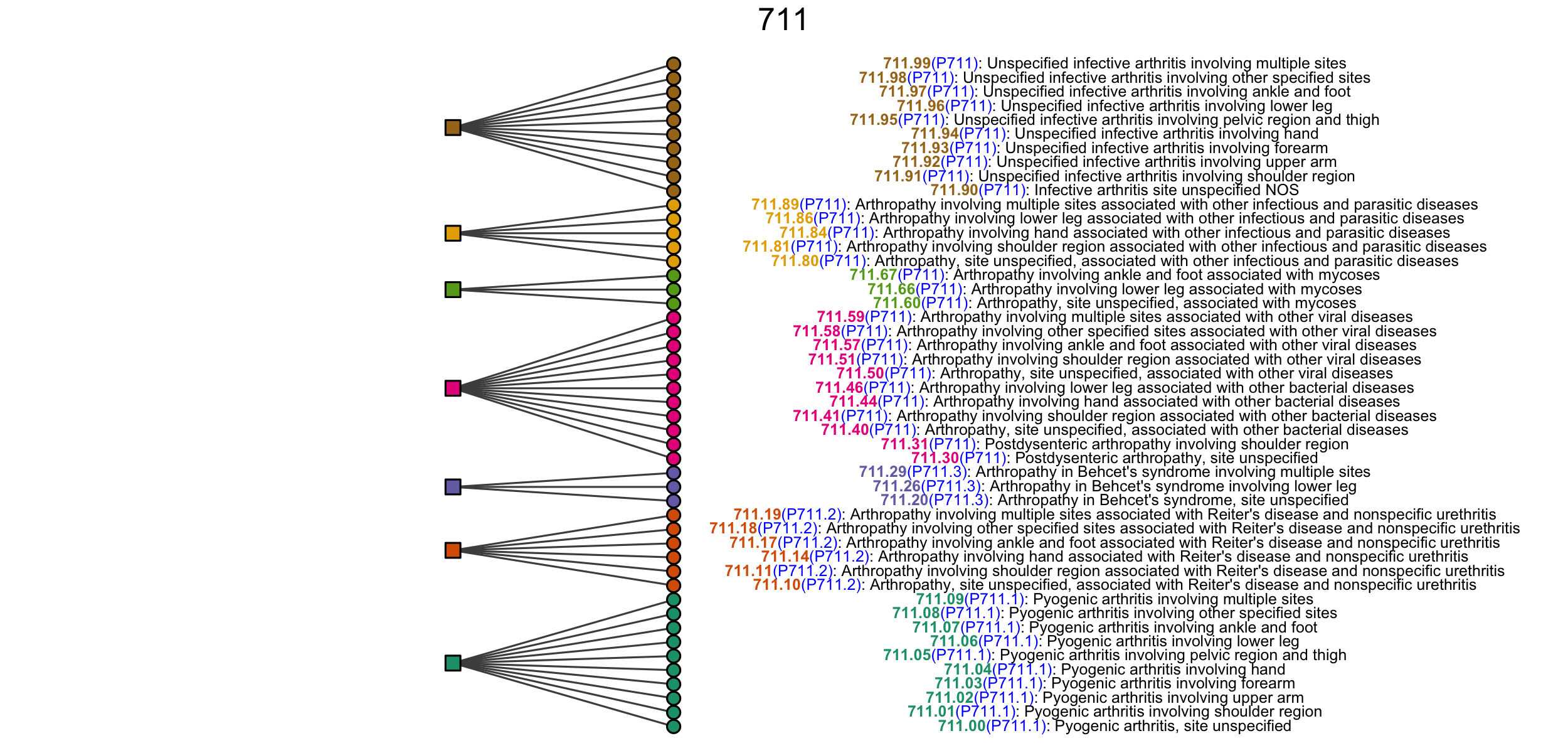}}\quad
        \subfloat[714]{\includegraphics[width=8cm,height=7cm]{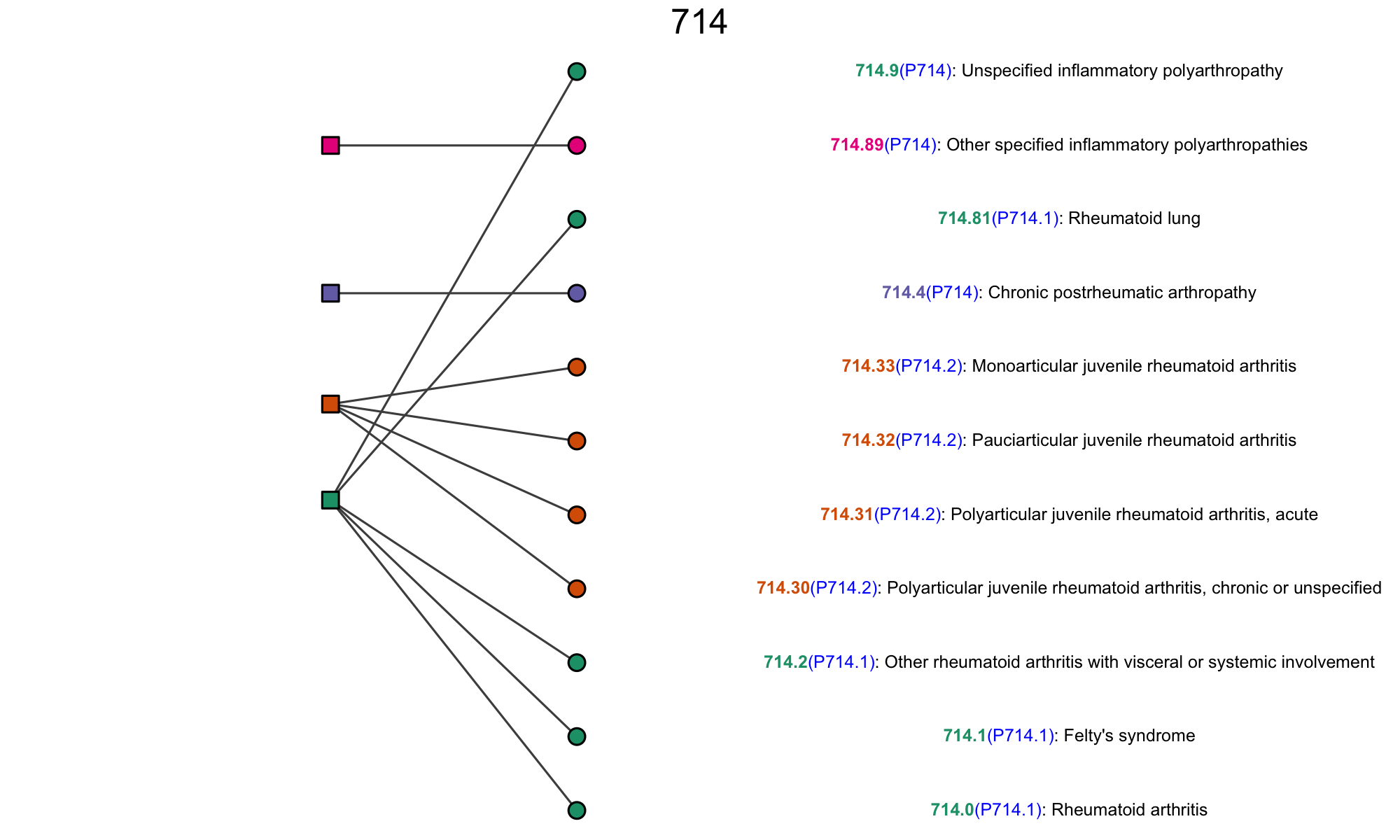}}
		\caption{Grouping result for arthropathies. Codes are colored by their membership. Codes in parenthesis colored in blue represent corresponding PheWAS codes.}
		\label{fig: 711&714}
	\end{center}
\end{figure}

To further demonstrate its efficacy, we zoom in to individual three-digit categories of ICD9 codes and examine their grouping structures compared to PheWAS. Figure \ref{fig: 359} shows a typical example that mvBSC$\subSNR$ based grouping result perfectly agrees with PheWAS.
In other cases, mvBSC$\subSNR$ turns out to be quite robust with only a few occasional mismatches to the best grid results. For example Figure \ref{fig: 368} compares the two corresponding results of category 368 in which only code 368.9 is grouped differently. Indeed mvBSC$\subSNR$ seems to be able to do a better job in this scenario in that unspecified visual disturbance is clinically similar to any other specified disturbances and thus is not necessarily parsed out. 
Figure \ref{fig: 711&714} (a) gives the clustering result on ICD9 codes starting with 711. PheWAS separates these codes into four groups, with a majority of the codes being grouped to represent {\em Arthropathy associated with infections} (P711), followed by {\em Pyogenic arthritis} (P711.1), {\em Reiter's disease} (P711.2), and {\em Behcet's syndrome} (P711.3). On the other hand, mvBSC$\subSNR$ separates these codes into seven concept groups with perfect agreement for codes in P711.1, P711.2 and P711.3. The main difference between mvBSC$\subSNR$ grouping and PheWAS grouping appears in codes that belong to P711 by PheWAS. While our method does not distinguish postdysenteric arthropathy from arthropathy associated with viral and bacterial diseases, it can perfectly set apart anthropathy associated with unspecified infective arthritis, other infectious and parasitic diseases, and mycoses.
As a final example, ICD9 codes starting with 714 consist of Rheumatoid Arthritis (RA) and Juvenile Rheumatoid Arthritis (JRA). Clinically, these are distinct diseases and thus should be grouped separately, despite adjacent coding representations and similar terminologies. Figure \ref{fig: 711&714} (b) well demonstrates that our approach is able to distinguish between these two conditions.

\section{Concluding remarks}
In this paper, we introduce a novel spectral clustering method that incorporates multiple data sources and leverages the prior distance knowledge among nodes. More specifically, the novelty consists of two main parts. First,  a consensus clustering is realized by the means of a weighted sum of membership-encoded matrices that attempts to drag all views to a common ground while allowing between-view heterogeneity. Second, the proposed approach effectively leverages the prior distance knowledge via the banding step. The statistical performance of the proposed method is thoroughly studied under a multi-view stochastic block model (mvSBM) framework. In particular, we demonstrate the effect of a banding operation on reducing the mean absolute operator-norm error bound to $ \max \left\{\left( n_{max}^{{1 \over 2\alpha + 1}}(\log n)^{{\alpha \over 2\alpha + 1}} \right), \sqrt{{\delta \over d_0} \log n },\log n \right\}$. Reducing to the standard result up to a $\sqrt{\log n}$ factor, this bound shows the robustness of our approach to the abscence of banding. In addition, banding by the distance also encourages a desired sparseness pattern in the observed similarity matrix and the sparseness level can be well controlled by the choice of $\delta$. Both simulations and the real data analysis demonstrate the effectiveness of the proposed mvBSC method to dramatically improve the clustering performance for a network with ordered nodes. We also provide a simple SNR based rule of choosing the weights that is intuitive and easy to follow in practice. However, we would like to make additional notes that this guideline may not yield a satisfactory result if a more complex hetereogeneity pattern is present in the data. In this paper, we focus on the case where heterogeneity is only allowed across different views. Relaxing the homoscedasticity assumption within each view warrants further research.  

\bibliography{paper_refs}

\appendix
\section{Appendix}

\subsection{Proof for Lemma \ref{lemma: F to H} }

\begin{proof}
	
	$\forall h >2\delta$, since $d(v_i,v_j) \leq d(v_i, v_{c_{g_i}}) + d(v_{c_{g_i}}, v_{c_{g_j}}) + d(v_j, v_{c_{g_j}}) \leq d(v_{c_{g_i}}, v_{c_{g_j}}) + 2\delta$, $d(v_i, v_j)>h$ implies $d(v_{c_{g_i}}, v_{c_{g_j}}) >h-2\delta>0$. Therefore, for any $i$,
	$\sum_j \{ |\Wscr^s_{ij}|: d(v_i,v_j) > h \} = \sum_j \{|\Omega^s_{g_i g_j}|: d(v_{c_{g_i}}, v_{c_{g_j}}) > h-2\delta\} \leq n_{max} \sum_l \{|\Omega^s_{g_il}|: d(v_{c_{g_i}}, v_{c_{g_j}}) > h-2\delta\} \leq n_{max}L            \left({h-2\delta \over d_0}\right)^{-\alpha_s}$. It is easy to see that $n_{min}\beta \leq \gamma_{K}(\Wscr^s) \leq  \gamma_1(\Wscr^s) \leq n_{max}/\beta$, which completes the proof.
\end{proof}

\subsection{Proof of Theorem \ref{thm: optimal bandwidth}}
\begin{proof}
	
	Let $\Ebb_{ij}$ denote the $n \times n$ indicator matrix whose $(i,j)$-th and $(j,i)$-th entry is 1 and 0 elsewhere, then
	\begin{eq}
		\label{eq: each W}
		B_{h}(\Wbb) - \Wscr &= B_{h}(\Wbb) - B_{h}(\Wscr) + B_{h}(\Wscr)  - \Wscr \\
		&= \left ( \sum_{ \underset{d(v_i,v_j)\leq h}{1 \leq i < j \leq n}} (\Wbb_{ij} - \Wscr_{ij}) \Ebb_{ij} \right ) + \diag\left(\omega_0 - \Wscr_{ii}\right)+ B_{h}(\Wscr)  - \Wscr
	\end{eq}
	Obviously, $\| (W_{ij} - \Wscr_{ij}) \Ebb_{ij} \| \leq 4L$, and $\{ (W_{ij} - \Wscr_{ij}) \Ebb_{ij} \}_{d(v_i,v_j)\leq h, 1 \leq i < j \leq n}$ is a sequence of independent random matrices, using the matrix Bernstein inequality given in Theorem 6.6.1 in \cite{tropp2015introduction},
	\begin{eq}
		\label{eq: matrix bernstein}
		\sE_{\Zbb^*} \norm{ B_{h}(\Wbb) - \Wscr } \leq \sqrt{2 \zeta \log n} + Ln_{max}(h-2\delta/d_0)^{-\alpha} + {4L \over 3} \log n + 2L
	\end{eq}
	where $$\zeta = \norm{ \sum_{ \underset{d(v_i,v_j)\leq h}{1 \leq i < j \leq n}} \sE_{\Zbb^*} \left[(W_{ij} - \Wscr_{ij}) \Ebb_{ij} \right ]^2} = 2\sigma^2 \lfloor h/ d_0\rfloor \leq 2\sigma^2 h/ d_0 .$$ 
	
	To find a proper order of $h$, it is easy to see that 
	\begin{eq}
		\sE_{\Zbb^*} \norm{ B_{h}(\Wbb) - \Wscr } & \leq 2 \sqrt{ \sigma^2 {h  \over d_0} \log n } + Ln_{max}(h-2\delta)^{-\alpha} + + {4L \over 3} \log n + 2L \\
		&\leq 2 \sqrt{ \sigma^2 {  h-2\delta \over d_0} \log n} + L n_{max} (h-2\delta/d_0)^{-\alpha} + 2 \sqrt{ \sigma^2 {  2\delta \over d_0} \log n}  + {4L \over 3} \log n + 2L
	\end{eq}
	Setting $\sqrt{{h-2\delta \over d_0} \log n} \asymp n_{max}(h - 2\delta/d_0)^{-\alpha}$ yields $\frac{h-2\delta}{d_0} \asymp \left( \frac{n_{max}}{\sqrt{\log n} } \right)^{\frac{2}{ 2\alpha + 1}} $, and 
	\begin{eq}
		\sE_{\Zbb^*} \norm{ B_{h}(\Wbb) - \Wscr } & \lesssim \max \left(\left( n_{max}^{{1 \over 2\alpha + 1}}(\log n)^{{\alpha \over 2\alpha + 1}} \right), \sqrt{{\delta \over d_0} \log n},\log n \right) 
	\end{eq}
\end{proof}	

\def\Ibb{\mathbb{I}}

\subsection{Proof of Theorem \ref{thm: mis-clustered error rate}}
\begin{proof}
	Recall that $(\Zbbhat, \Abbhat ) = \argmin_{\Zbb \in \Zscr_{n,K}, \Abb \in \Rcal^{K \times K}} \norm{\Ubbhat^*_{\blambda} - \Zbb \Abb}_F^2 $, thus
	$$\|\Zbbhat \Abbhat- \Ubb^*Q\|_F^2 \leq 2\|\Zbbhat \Abbhat -  \Ubbhat_{\blambda}^*\|^2_F + 2\|\Ubbhat_{\blambda}^* - \Ubb^*Q\|^2_F \leq 4\|\Ubbhat_{\blambda}^* - \Ubb^*Q\|^2_F $$
	%
	It follows from the results from \cite{lei2015consistency} that
	\begin{eq}
		\label{eq: lei} 		
		&\| \Ubbhat^*_{\blambda} - \Ubb^* Q\|^2_F \leq 8K \| \sum_{s=1}^m \lambda_s \Ubbhat^s \Ubbhat^{s\trans}- \Ubb^* \Ubb^{* \trans}\|^2 \leq  8mK \sum_{s=1}^m \lambda^2_s \|\Ubbhat^s\Ubbhat^{s\trans} - \Ubb^s \Ubb^{s \trans}\|^2  \\
		\mbox{and} \quad& \| (\Ibb -  \Ubbhat^s\Ubbhat^{s \trans}) \Ubb^s\Ubb^{s \trans}  \|  \leq 2 {\|B_{h_s}(\Wbb^s) - \Wscr^s\| \over \gamma^s_{n,K}}
	\end{eq}
	where $\Ibb$ is the identity matrix. 

	In addition, since 
	$\| (\Ibb -  \Ubbhat^s\Ubbhat^{s \trans}) \Ubb^s\Ubb^{s \trans}  \| = \|\Ubbhat^s\Ubbhat^{s\trans} - \Ubb^s \Ubb^{s \trans}\|  $,
	we have
	\begin{eq}
		\label{eq: squared deterministic upper bound}
		\| \Ubbhat^*_{\blambda} - \Ubb^* Q\|^2_F
		\leq  32mK  \sum_{s=1}^m   \left(\frac{\lambda_s}{\gamma^{s}_{n,K}} \right)^2 \norm{B_{h_s}(\Wbb^s)-\Wscr^s}^2 .
	\end{eq}
	It follows that
	\begin{eq}
		|\Mcal_{\blambda}| = \sum_{i \in \Mcal_{\blambda}} 1 
		& \leq 2n_{max} \sum_{i\in \Mcal_{\blambda}} \|\Abbhat_{g_i \cdot} - U_{i \cdot}^*Q\|^2_2 
		\leq 2n_{max} \sum_{i=1}^n \|\Abbhat_{i\cdot} - \Ubb_{i\cdot}^*Q\|^2_2 \\
		& = 2n_{max} \|\Zbbhat \Abbhat - \Ubb^*Q\|_F^2 
		\leq 8n_{max} \|\Ubbhat^*_{\blambda} - \Ubb^*Q\|_F^2 \\
		& \leq 256 mn_{max}K \sum_{s=1}^m   \left( \frac{\lambda_s}{\gamma^{s}_{n,K}}\right)^2 \norm{B_{h_s}(\Wbb^s)-\Wscr^s}^2
	\end{eq}
	
	Using Bernstein inequality given in Theorem 6.6.1 in \cite{tropp2015introduction}, for all $t \geq 0$,
	\begin{eq}
		\Pr(\norm{ B_{h_s}(\Wbb^s) - \Wscr^s } \geq t) & \leq n\exp\left \{ \frac{-t^2/2}{ 2\sigma_s^2 h_s/ d_0 + 4Lt/3} \right\} 
		\leq \exp\left \{ \log n - \frac{t^2/4}{ b_s( h_s/d_0 + t) } \right\} \\
		& \leq \left\{ \begin{array}{cc}
			\exp\left \{ \log n - \frac{t^2}{ 8b_s h_s/d_0 } \right\} & \mbox{if } t \leq h_s/d_0 \\
			\exp\left \{ \log n - \frac{t}{ 8b_s } \right\} & \mbox{if } t \geq h_s/d_0 
		\end{array} \right .
	\end{eq} 
	where $b_s = \max(\sigma_s^2,2L/3)$. Hence, for any $r > 0$, with probability at least $1-n^{-r}$, if $n_{max} \gtrsim (\log n)^{\alpha_s + 1} $
	\begin{eq}
		\label{eq: prob bound 1}
		\norm{ B_{h_s}(\Wbb^s) - \Wscr^s }  \leq \left\{ \begin{array}{cc}
			c_s(r+1) \left( n_{max}^{{1 \over 2\alpha_s + 1}}(\log n)^{{\alpha_s \over 2\alpha_s + 1}} \right) & \mbox{if } \; {\delta \over d_0} = o\left(  \left( \frac{n_{max}}{\sqrt{\log n} } \right)^{\frac{2}{ 2\alpha_s + 1}} \right) \\
			c_s(r+1) \sqrt{\frac{\delta}{d_0} \log n }& \mbox{otherwise }   
		\end{array} \right .
	\end{eq} and if $n_{max} = o((\log n)^{\alpha_s+1})$,
	\begin{eq}
		\label{eq: prob bound 2}
		\norm{ B_{h_s}(\Wbb^s) - \Wscr^s }  \leq \left\{ \begin{array}{cc}
			c_s(r+1) \sqrt{\frac{\delta}{d_0} \log n }& \mbox{if } \; {\delta \over d_0} \gtrsim  \sqrt{\log n}  \\
			c_s(r+1) \log n & \mbox{otherwise }   
		\end{array} \right .
	\end{eq} 
	for some positive constant $c_s$ that depends on $b_s$.
	Applying the union bound,  for any $r > 0$, with probability at least $1-mn^{-r}$, 
	\begin{eq}
		\label{eq: squared deterministic upper bound}
		\| \Ubbhat^*_{\blambda} - \Ubb^* Q\|^2_F
		\leq  32mK(r+1)^2  \sum_{s=1}^m  \left( \frac{\lambda_s c_s}{\gamma^{s}_{n,K}}\right)^2 \max \left( n_{max}^{{2 \over 2\alpha_s + 1}}(\log n)^{{2\alpha_s \over 2\alpha_s + 1}} , {\delta \log n\over d_0},(\log n)^2 \right)
	\end{eq} 
	
	Therefore, dropping some constant terms not involving with $n$, with probability at least $1-m/n$,
	\begin{eq}
		\label{eq: general err bound}
		\frac{|\Mcal|}{n} = O_p\left( {n_{max} \over n} \sum_{s=1}^m \left( \frac{\lambda_s}{\gamma^{s}_{n,K}}\right)^2 \max \left( n_{max}^{{2 \over 2\alpha_s + 1}}(\log n)^{{2\alpha_s \over 2\alpha_s + 1}} ,{\delta \log n\over d_0}, (\log n)^2 \right) \right)
	\end{eq}	
\end{proof}

\subsection{Proof of Corollary \ref{cor: special case}}
\begin{proof}
	Recall that $\gamma^s_{n,K}$ is the K-th largest eigenvalue of $\Wscr^s$, from Lemma \ref{lemma: F to H}, $\gamma_{n,K}^s$ is at least at the scale of $n_{min}$. The results are natural simplifications of (\ref{eq: general err bound}).
\end{proof}

\subsection{Proof of Theorem \ref{thm: q_lambda} }
\begin{proof}
	Recall that
	\begin{eq}
		\label{eq: lei} 		
		\norm{ \sum_{s=1}^m \lambda_s \Ubbhat^s \Ubbhat^{s\trans}- \Ubb^* \Ubb^{* \trans}}^2 &\leq  m  \sum_{s=1}^m \lambda^2_s \|\Ubbhat^s\Ubbhat^{s\trans} - \Ubb^s \Ubb^{s \trans}\|^2  \\
		& \leq 4m \sum_{s=1}^m \left(\frac{\lambda_s}{\gamma^s_{n,K}}\right)^2 \|B_{h_s}(\Wbb^s) - \Wscr^s\|^2 
	\end{eq}
	
	From (\ref{eq: each W}), 
	\begin{eq}
		\label{eq: upper bound on MSE}
		\norm{ B_{h_s}(\Wbb^s) - \Wscr^s} ^2\leq 
		3(2L)^2 + 3\left(Ln_{max} (h_s-2\delta / d_0)^{-\alpha_s} \right)^2 + 3 \norm{ \sum_{ \underset{d(v_i,v_j)\leq h_s}{1 \leq i < j \leq n}}  (\Wbb^s_{ij} - \Wscr^s_{ij}) \Ebb_{ij} }^2
	\end{eq}
	
	Using Matrix second-moment inequality given in \cite{chen2012masked},
	\begin{eq}
		\label{eq: berinstein on MSE}
		\sE_{\Zbb^*} \norm{ \sum_{ \underset{d(v_i,v_j) \leq h_s}{1 \leq i < j \leq n}} (\Wbb^s_{ij} - \Wscr^s_{ij}) \Ebb_{ij} }^2   &\leq \left(  2\sqrt{e\log n} \sqrt{ \sigma_s^2 h_s / d_0}  + 4e\sigma_s  \log n   \right)^2 \\	
		&= 4\sigma_s^2 e\log n \left(  \sqrt{ h_s / d_0}  + 2\sqrt{e \log n}    \right)^2 \\
		&\leq 8\sigma_s^2 \left(  h_s / d_0 + 4e \log n \right) e\log n 
	\end{eq}
	
	Setting $2 \sqrt{{h_s-2\delta \over d_0} \log n} = L n_{max} \left( {h_s-2\delta \over d_0} \right)^{-\alpha_s} $,
	\begin{eq}
		\sE_{\Zbb^*} \norm{ B_{h_s}(\Wbb^s) - \Wscr^s }^2 \leq 12L^2 + 12 {h_s \over d_0} (1 + 2e\sigma_s^2) \log n   + 96 \sigma_s^2e^2(\log n)^2	
	\end{eq}
	and 
	\begin{eq*}
		\sum_{s=1}^m \left(\frac{\lambda_s}{\gamma^s_{n,K}}\right)^2 \|B_{h_s}(\Wbb^s) - \Wscr^s\|^2 &\leq {12 \log n \over d_0} \sum_{s=1}^m \left(\frac{\lambda_s}{\gamma^s_{n,K}}\right)^2 (1 + 2e\sigma_s^2 ) h_s + 96 (e\log n)^2 \sum_{s=1}^m \left(\frac{\sigma_s \lambda_s}{\gamma^s_{n,K}}\right)^2 \\
		& + 12L^2 \sum_{s=1}^m \left(\frac{\lambda_s}{\gamma^s_{n,K}}\right)^2 
	\end{eq*}
	Since $n_{max} \gtrsim (\log n)^{\alpha_s+1}$, $h_s/ d_0 \gtrsim \log n, s=1,...,m$, dropping some negligible and constant terms, we have
	\begin{eq*}
		\sE_{\Zbb^*} \norm{ B_{h_s}(\Wbb^s) - \Wscr^s }^2 \leq C h_s \sigma_s^2 \log n
	\end{eq*} and
	\begin{eq*}
		\sum_{s=1}^m \left(\frac{\lambda_s}{\gamma^s_{n,K}}\right)^2 \sE_{\Zbb^*} \|B_{h_s}(\Wbb^s) - \Wscr^s\|^2 &\leq C \log n \sum_{s=1}^m \left(\frac{\sigma_s \lambda_s}{\gamma^s_{n,K}}\right)^2 h_s 
	\end{eq*}

\end{proof}

\end{document}